\documentclass[12pt]{article}
 \input epsf.sty

\usepackage{amsmath,amsfonts}

\newdimen\hoogte    \hoogte=18pt    
\newdimen\breedte   \breedte=20pt   
\newdimen\dikte     \dikte=0.5pt    
\def\beginYoung{
       \begingroup
       \def\vr{\vrule height0.8\hoogte width\dikte depth 0.2\hoogte}
       \def\fbox##1{\vbox{\offinterlineskip
                    \hrule height\dikte
                    \hbox to \breedte{\vr\hfill##1\hfill\vr}
                    \hrule height\dikte}}
       \vbox\bgroup \offinterlineskip \tabskip=-\dikte \lineskip=-\dikte
            \halign\bgroup &\fbox{##\unskip}\unskip  \crcr }
\def\End@Young{\egroup\egroup\endgroup}
\newenvironment{Young}{\beginYoung}{\End@Young}

\newcommand{\bea}{\begin{eqnarray}}
\newcommand{\eea}{\end{eqnarray}}
\newcommand{\bean}{\begin{eqnarray*}}
\newcommand{\eean}{\end{eqnarray*}}

\def\O #1{\overline{#1}}

\def\W #1{\widetilde{#1}}

\def\ba{\begin{array}}
\def\ea{\end{array}}
\def\beq{\begin{equation}}
\def\eeq{\end{equation}}
\def\braket#1{\left\langle #1 \right\rangle}
\def\bra#1{\left\langle #1\right|}
\def\ket#1{\left| #1\right\rangle}

\def\Tr{\mathop{\rm Tr}}
\def\det{\mathop{\rm det}}
\def\det{\mathop{\rm det}}
\def\sign{\mathop{\rm sign}}
\def\diag{\mathop{\rm diag}}

\def\to{\rightarrow}

\def\a{{\alpha}}
\def\b{{\beta}}

\newcommand{\eqn}{\begin{eqnarray}}
\newcommand{\enn}{\end{eqnarray}}

\def\a{\alpha}
\def\b{\beta}

\def\Label#1{\label{#1}%
  \smash{\hbox to0pt{\raise1ex\hbox{\tiny[#1]}\hss}}}
\def\noLabels{\let\Label=\label}
\def\nobbibitem{\let\bbibitem=\bibitem}

\def\ket{\rangle}
\def\bra{\langle}

\def\CO{{\cal O}}

\newcommand{\bbibitem}[1]{\bibitem{#1}\marginpar{#1}}

\newcommand{\ads}[1]{{\rm AdS}_{#1}}

\newcommand{\be}{\begin{equation}}
\newcommand{\ee}{\end{equation}}
\def\tr{\hbox{tr}}

\newcommand{\CN}{{\cal N}=4}

\begin{document}

\renewcommand{\thepage}{\arabic{page}}
\setcounter{page}{1}
\noLabels 
\nobbibitem 

\rightline{hep-th/0411205}
\rightline{UPR-1095-T, MAD-TH-04-12}
\rightline{NSF-KITP-04-123}
\vskip 1cm
\centerline{\Large \bf 
D-branes in Yang-Mills theory and}
\vskip 0.5cm
\centerline{\Large \bf Emergent Gauge Symmetry}
\vskip 1cm

\renewcommand{\thefootnote}{\fnsymbol{footnote}}
\centerline{{\bf Vijay Balasubramanian}$^{1}$\footnote{vijay@physics.upenn.edu},
{\bf David Berenstein}$^{2,3}$\footnote{dberens@physics.ucsb.edu},
{\bf Bo Feng}$^{4}$\footnote{fengb@ias.edu},
{\bf Min-xin Huang}$^{5}$\footnote{minxin@physics.wisc.edu}
}

\centerline{${}^1$\it David Rittenhouse Laboratories, University of
Pennsylvania}
\centerline{\it Philadelphia, PA 19104, U.S.A.}

\centerline{${}^2$\it  Department of Physics, UCSB,}
\centerline{\it Santa Barbara, CA 93106, USA.} 

\centerline{${}^3$\it  Kavli Institute for Theoretical Physics}
\centerline{\it Santa Barbara, CA 93106, USA.} 

\centerline{${}^4$\it  Institute for Advanced Study}
\centerline{\it Olden Lane, Princeton, NJ 08540, USA.}

\centerline{${}^5$\it  Department of Physics, University of Wisconsin}
\centerline{\it Madison, WI 53706, USA.}

\setcounter{footnote}{0}

\renewcommand{\thefootnote}{\arabic{footnote}}

\begin{abstract}
Four-dimensional supersymmetric SU(N) Yang-Mills theory on a sphere has highly charged baryon-like states built from anti-symmetric combinations of the adjoint scalars.  We show that these states, which are equivalently described as holes in a free fermi sea of a reduced matrix model, are D-branes.  Their excitations are stringlike and effectively realize Dirichlet and Neumann boundary conditions in various directions.   The low energy brane dynamics  should realize an emergent gauge theory that is local on a new space.   We show that the Gauss' Law associated to this emergent gauge symmetry appears from combinatorial identities relating the stringy excitations.   Although these excitations are not BPS, they can be near-BPS and we can hope to study them in perturbation theory.  Accordingly, we show that the Chan-Paton factors expected for strings propagating on  multiple branes arise dynamically,  allowing the emergent gauge symmetry to be non-Abelian.    

\end{abstract}

\section{Introduction}

It is known that Yang-Mills theory with many colours has a relation to a theory of closed strings \cite{thooft}.  Indeed, the AdS/CFT correspondence showed that certain Yang-Mills theories are exactly equivalent to fundamental string theories in AdS backgrounds \cite{jthroat}.   Since such closed string theories also contain D-branes and hence open string dynamics, one expects that Yang-Mills theories contains sectors which display open string dynamics and thus the dynamical emergence of new non-Abelian gauge symmetries.  The purpose of this paper is to study how this happens in the $\CN$ SU(N) Yang-Mills theory on a sphere that is dual to IIB string theory in global $\ads{5} \times S^5$.

Our computational techniques will require extrapolation from weak to strong coupling, and thus we are particularly interested in states that do not receive large quantum corrections.  For this reason it is convenient to start with 1/2-BPS D-brane states of $\CN$ Yang-Mills theory, which we describe in Sec.~2.    The dual description of these states in string theory on $\ads{5} \times S^5$ is as ``giant gravitons'' \cite{MST,akietal,goliath}, or spherical branes, that expand out into the $\ads{5}$ or into the $S^5$.  The relevant Yang-Mills operators were described in \cite{BBNS,CJR}.  Extensive study of these states has culminated with the understanding that there is a reduced matrix model which describes them. The matrix model is studied most efficiently in terms of the free fermionic description of the eigenvalues \cite{Btoy}, and it turns out that the two types of giant gravitons can be understood either as the fermions themselves, or as hole wavefunctions.   This insight has been further confirmed by the construction of all half BPS supergravity solutions with asymptotic $AdS_5\times S^5$ boundary conditions \cite{linluninmalda} (see also \cite{Calda} for some earlier work).

One class of these states -- corresponding to spherical D-branes that expand into $\ads{5}$ -- is described in the semiclassical limit as the a time-dependent analog of  a Coulomb branch for the field theory on a sphere \cite{akietal}.   We explain this in Sec.~3 and show therefore that the gauge dynamics of D-branes in spacetime is simply embedded in the residual gauge symmetry that survives the (time-dependent) Higgs effect in the field theory.   In Sec.~4 we describe the Yang-Mills operators dual to spherical branes that expand into $S^5$ and establish that the excitations around states created by these operators are open string-like and realize the Dirichlet and Neumann boundary conditions expected for a D-brane.   This completes the evidence accumulated in \cite{BHLN,Bgiant,AABF} for the branelike nature of such states.  In the M-theory plane wave it has been argued that tiny giant gravitons are the partons that give rise to the matrix degrees of freedom of the plane wave matrix 
model \cite{ShJ}. One can extrapolate this idea further and say that giant gravitons generate the global $AdS_5\times S^5$ geometry \cite{linluninmalda}, where we have replaced them by the flux background they generate.


 In Sec.~5 we propose operators describing the open string excitations of multi-brane states. They  are not naive products of single brane operators, but are strongly constrained by the required symmetries.
The low energy dynamics of such strings should realize a new gauge theory on the brane worldvolume.   The simplest manifestation of a gauge theory on a compact space is that Gauss' Law only allows excitations within vanishing total charge.  In our proposal, Gauss' law emerges from combinatorial identities due to which states that are gauge-invariant in the original Yang-Mills theory, but charged in the emergent theory on the D-branes, are impossible to construct.\footnote{Previous work by Sadri and Sheikh-Jabbari~\cite{sadri} has shown that  strings stretching between spherical branes in AdS can be realized as solutions to the Born-Infeld action on these branes.    Gauss' law emerges in this spacetime picture in the construction of consistent solutions to the equations of motion on a compact space.   Since this compact space is not directly part of the manifold on which the Yang-Mills theory is defined, it is very interesting to see how Gauss' emerges from the dual field theory perspective.} While such open string excitations on D-branes are not BPS, they can be near-BPS and hence one can hope to study their dynamics in perturbation at weak coupling and then extrapolate to strong coupling.  
 
Carrying out such studies in Sec.~6, we show that the Chan-Paton factors that appear in open-string loop diagrams in the presence of multiple branes emerge dynamically from Yang-Mills calculations.  In particular, we find that as two D-branes separate, the associated Chan-Paton factor interpolates between 2 and 1.     The section concludes with some attempts to study the interactions of the open strings that have emerged from Yang-Mills theory.

\section{$U(N)$ Matrix models and 1/2-BPS states}

We are interested in the 1/2-BPS states of $\CN$ Yang-Mills theory that are dual to compact D-branes (giant gravitons) in $\ads{5} \times S^5$.   To construct the relevant operators it is helpful to recall some elementary facts about the 1/2-BPS operators.    

Recall that $\CN $ Yang-Mills theory has three chiral superfields.  The 1/2-BPS operators are constructed as gauge invariant combinations of any one of these fields, say $\phi$, and will be BPS with respect to an R-symmetry,  $J$, which generates an $SO(2) \subset SO(6)$.    If  $\Delta$ is the generator associated to dilatations on the Euclidean plane ($R^4$), the BPS bound is  
\begin{equation}
\Delta=J \, .
\end{equation}  
Anti-BPS operators, constructed from $\bar\phi$ satisfy $\Delta = -J$.   Given an operator on $R^4$, the state-operator correspondence of the CFT can be used to give a state of the Lorentzian field theory on $S^3 \times R$.      The operator $\Delta$ becomes Hamiltonian for the latter.  Specifically, given an operator $\CO$ in the Euclidean theory, we apply it to the origin, i.e. $\CO(0)$, and then conformally map and then  Wick rotate  to Lorentzian $S^3 \times R$.   We will now argue that all the 1/2-BPS states of the theory on $S^3 \times R$ can described in the S-wave reduction of the Yang-Mills theory to a matrix quantum mechanics.

The scalar field $\phi(x)$ has a scaling dimension $\Delta = 1$.  The Taylor expansion of this field on a surface of fixed time $t$ is
\begin{equation}
\phi(x) = \phi(0) +
\sum_{k=1}^\infty {1 \over k!} \left[\partial_{\mu_1} \cdots \partial_{\mu_k} \phi\right]_{x=0} x^{\mu_1} \cdots x^{\mu_k}
\Label{Taylorexp}
\end{equation}
where the $x^\mu$ are spatial coordinates.  Because  derivatives commute, the expansion is symmetrized over $x^\mu$, and so  this is also an expansion in spherical harmonics (up to traces).   Since a derivative $\partial_\mu$ has scaling dimension $1$,
the ${\rm m}^{{\rm th}}$ term in the Taylor expansion (or equivalently in a decomposition in spherical harmonics) 
has a scaling dimension $\Delta = m+1$.    However, the R-charge of all terms in the expansion is still $J = 1$.    Thus, the 1/2-BPS states, which must have $\Delta = J$, are all constructed out of the S-wave reduction of $\phi$, which we will write as $\phi_{(0)}$.

For example, a 1/2-BPS state corresponding to a supergravity mode with angular momentum $k$ on the $S^5$ part of the bulk geometry would be given by
\begin{equation}
\tr((\phi^\dagger_{0)})^k)|0\rangle \, .
\end{equation}
Similarly, to construct all 1/2-BPS states of $\CN$ Yang-Mills on $S^3$, we need only the S-wave parts of the spherical harmonic decompositions of all the scalar fields.  Thus the study of 1/2-BPS operators relative to a particular $U(1)$ R-symmetry generator $J$  reduces to the study of  the quantum mechanics of the mode $\phi_{0}$:  a one matrix quantum mechanics.

We must also decompose the gauge fields of the Yang-Mills theory into spherical harmonics.  Because of the vector index, only the S-wave of the timelike component $A_0$ has a coupling to $\phi_{(0)}$.   This leaves a $U(N)$ gauged matrix quantum mechanics.

At this point, one could write the effective dynamics of this mode integrating out everything else in the  field theory.  In the free field limit we simply get a matrix harmonic oscillator. Due to supersymmetry, we expect that this will not change as we increase the coupling constant. Basically, we expect fermions and bosons to cancel and prevent the appearance of an additional potential for the mode $\phi_{0,0}$, and we expect to be able to extrapolate weak coupling results to strong coupling for certain states which are 'close' to these configurations. 

The half BPS operators of $\CN$ Yang-Mills theory will in be one to one correspondence with the states
of this gauged matrix quantum mechanics.   Now the problem is to describe a complete basis for these states. As shown in \cite{CJR}, a complete set of such states is given by Schur polynomials in the raising operators. These are constructed as characters of group elements 
in the different irreducible representations of $U(N)$.
An irreducible representation of $U(N)$ built out of $m$ copies of the fundamental representation $V$ is correlated with a representation of the symmetric group $S_m$.   As a result,  it can be classified by a Young tableaux with $m$ boxes  
\footnote{Some subtleties that arise if the group is $SU(N)$ rather than $U(N)$ have been explained recently
in the paper \cite{DeM}.   In the end, one can write the theory for the $U(N)$ group and notice that the $U(1)$ degrees of freedom
factorize.  One can proceed by  doing the full analysis for $U(N)$ and afterwards remove the unwanted degrees of freedom.   These issues are subleading in the $1/N$ expansion, and hence we will ignore them in our analysis.}

For example, a totally antisymmetric representation of $U(N)$ maps to  a Young tableau with one column (the length of which cannot exceed $N$).    A totally symmetric representation of $U(N)$ maps to a Young tableau with one row.  These two types of representations have been conjectured  to describe the Yang-Mills operators dual to spherical D-branes  that expand into $S^5$ (giant gravitons) and $AdS_5$ (``dual'' giant gravitons) respectively \cite{BBNS,CJR}.    It has been established that one can correctly reproduce the spectrum of fluctuations of the DBI action for a single maximal size giant graviton  \cite{Bgiant}. Also, the operators dual to a single maximal giant graviton with some number of attached strings in the plane wave limit have been described in \cite{BHLN} although it has not been shown that the Yang-Mills theory dynamically imposes  the Dirichlet and Neumann boundary conditions that appear for open strings in the bulk spacetime description.   (We will demonstrate that this indeed happens in Sec.~\ref{baryonsasbranes}.)

Given that 1/2-BPS branes can be described as states in a gauged one matrix model, we can seek a description in terms of eigenvalues of the matrix $\phi_{(0)}$ (these are gauge invariant up to permutations) which effectively form a theory of free fermions in the harmonic oscillator. 
It turns out that the Young tableaux basis and the basis of Slater determinants of harmonic oscillator 
wave functions coincide. 
As shown in \cite{Btoy}, there is a beautiful picture in which giant gravitons of energy  $m$ (expanding on $S^5$ arise as a single column Young tableaux with $m$ boxes, and as holes in the Fermi sea of energy $m$.   Conversely, dual giant gravitons of energy $m$ (expanding into $AdS_5$) arise as a single row Young tableaux with $m$ boxes, and as  energy $m$ excitations of a single eigenvalue over the top of the Fermi sea. This is best described in the phase space plane of the eigenvalues, where
large collections of fermions put together will form droplets of an incompressible fluid.
This same behavior  has been seen also in the full supergravity solutions associated to 
the fermion droplet configurations \cite{linluninmalda}.

One of the key salient features of D-branes is that they give rise to matrix-valued degrees of freedom \cite{matrix}:
it is well-known that the open strings on D-branes realize a gauge theory on the brane worldvolume at low energies.   Since the Yang-Mills operators described above are dual to these D-branes, they must also realize this gauge theory if the AdS/CFT correspondence is valid.   How and whether this happens is an important question that we seek to answer.  The branes arising from the antisymmetric representations (which are dual to giant gravitons expanded into $S^5$) are the nearest analogue of baryons in $\CN$ Yang-Mills.  Hence it is extremely interesting to establish whether and how the low-energy dynamics around such baryon-like states realizes a new gauge theory with gauge group unrelated to that of the original Yang-Mills theory.

\section{The ``Coulomb branch'' on a sphere}

In the next section we will study the emergence of new gauge symmetries from the dynamics around baryon-like states, but it is useful to first examine a simpler realization of D-branes within Yang-Mills theory.  This is simply the familiar physics of the Coulomb branch.   In the present case, the novelty is that the gauge theory is on a compact space, so there is no conventional Coulomb branch.  However,  there is an appropriate analogue \cite{akietal}, the physics of which describes spherical D-branes expanded into $\ads{5}$.    As argued above, the relevant states can be written in terms of the matrix degrees of freedom of one complex scalar of the Yang-Mills theory.  However, to understand the  fluctuations of the D-brane we must embed these matrix states into the full $\CN$ SYM theory.

The SYM action for the $SU(N)$ theory  on $S^3$ is given by 
\begin{eqnarray}
\int dt \int_{S^3} d vol \frac{N}{4\pi g^2 N} [\frac 12 (D\phi^i)(D \phi^i)+\frac 14([\phi,\phi])^2+
\frac12\phi^i\phi^i \nonumber \\
 +~\hbox{fermions and gauge fields}]
\end{eqnarray}
The normalization is chosen such that the gauge curvature does not depend on the coupling.  The mass terms of the bosons arises from $R\phi^2$ term arising from conformal coupling to the metric of $S^3$,   The 't Hooft coupling is $g^2N = \lambda$.

Consider a spherical D-brane expanding into $\ads{5}$ with a radius of order the AdS scale.    Such a brane is stabilized by its angular momentum \cite{MST,akietal,goliath}, which we write as $J=Nj$, with $j$ a $O(1)$ constant.   As described in \cite{Btoy} and above, such state is created  by giving a total energy $Nj$ to a single eigenvalue of the S-wave of a complex scalar field, say $\phi=(\phi^1+i\phi^2)/{\sqrt2}$.     The semiclassical description of such a state is
\begin{equation}
\phi(x) = \diag(A,0,0,\dots 0)e^{-i t}
\Label{semi1}
\end{equation}
exactly as previously proposed by \cite{akietal}.  Matching the energy $jN = \int_{S^3} d vol \frac{N}{4\pi \lambda} |A|^2$ gives 
us the semiclassical amplitude for the field
\begin{equation}
|A|^2 = \frac{4\pi \lambda j}{{\rm Vol}(S^3)}
\end{equation}
Since this depends only the 't Hooft coupling, such states are well defined in the large $N$ 't Hooft limit.

Expanding around this semiclassical state, the n$^{{\rm th}}$ spherical harmonic (i.e. the $(n/2,n/2)$ representation) of off-diagonal terms of  $\phi^i$ for $i=3,\dots 6$ wil have a mass:
\begin{equation}
m = \sqrt{1+ n+  \frac{4\pi \lambda j}{{\rm Vol}(S^3)}}
\, .
\Label{massivemass}
\end{equation} 
This is the energy of the state on the cylinder -- on the plane this would map to the conformal dimension. This is because the energy operator on $S^3$ is related to the radial quantization of the Euclidean field theory on $R^4$.   Thus, the effective description of the spherical D-brane in $\ads{5}$ in the CFT is as a (time dependent) Higgs mechanism which breaks the gauge group $SU(N)$ to $SU(N-1)\times U(1)$.

By similar reasoning we can see the dual Yang-Mills realization of the enhanced gauge symmetry on coincident D-branes in $\ads{5}$.   To describe two D-branes in AdS, following \cite{Btoy} we must excite two eigenvalues:
\begin{equation}
\phi\sim \begin{pmatrix} A_1&0&0&\dots\\
0&A_2&0&\dots\\
0&0&0&\dots\\
\vdots&\vdots&\vdots&\ddots
\end{pmatrix}
\end{equation}
breaking the symmetry down to $U(1) \times U(1) \times SU(N-2)$
The branes are coincident in this semiclassical description if $A_1 = A_2$ so that $U(1) \times U(1)$ is enhanced $U(2)$.   The phase of the eigenvalue is associated to the position of the brane along the circle on the bulk $S^5$ along which the brane is moving.  
 Thus, in this simple example, the gauge symmetry arising from D-branes in $\ads{5}$ is embedded within the original gauge symmetrty of the dual and is local in the original $S^3$ on which the latter is defined.   For example, Gauss' law on the D-brane is implemented trivially  because the $SU(N)$ theory dual to $\ads{5}$ implements Gauss' law.   
 
\paragraph{Vibrational spectrum of $AdS_5$ branes: }   The fluctuation spectrum of the massless modes of spherical branes in $\ads{5}$ was computed in \cite{Das}.  Surprisingly, the spectrum turns out to be independent of the  angular momentum and size of the brane.  For example, translating into Yang-Mills variables, the fluctuations of the transverse scalars of the brane in spin $(n/2,n/2)$ spherical harmonic should be reperesented by Yang-Mills operators have a conformal dimension with conformal dimension $n+1$.  Observe that at large 't Hooft coupling  $\lambda$ the Yang-Mills configuration corresponding to a single brane on $\ads{5}$ (\ref{semi1}) involves the Coulomb branch with gauge group broken to $SU(N-1) \times U(1)$.    As described above, at large $\lambda$, the $U(1)$ is decoupled from $SU(N-1)$  since all off-diagonal terms are very massive, and all low energy states can be decomposed under these groups.   The transverse fluctuations of the brane in AdS are represented in the Yang-Mills theory by modes of the scalars $\phi^i$ for $i=3,\dots 6$ that are neutral under both  $U(1)$ and  $SU(N-1)$ -- namely, these are diagonal fluctuations of the form
\begin{equation}
\phi^i \sim \begin{pmatrix}
a^\alpha & 0 & \cdots \\
0&0 & \cdots \\
0 & 0 & \cdots \\
\vdots & \vdots & \ddots
\end{pmatrix}
\end{equation}
The spatial $SO(4)$ rotation group of the field theory maps onto the $SO(4)$ rotation group of the brane in the bulk.   Decomposing the $\phi^i$ into spherical harmonics, the spin $(n/2,n/2)$ representation will then have dimension $(n+1)$, matching the bulk D-brane results \cite{Das}.  
 Supersymmetry then generates the remainder of the spectrum.   We might worry that the free field result might experience large quantum corrections, because these fluctuations are not  BPS. However, one can argue that the states in question arise as would-be goldstone modes associated to the 
spontaneous breaking of the $SO(6)$ symmetry and therefore their masses are protected. This is similar to arguments given in \cite{BMN} to reproduce the spectrum of the string on the pp-wave. In both cases, if one takes the Hamiltonian as 
$H_{eff}= \Delta-J$, the Hamiltonian is no longer $SO(6)$ symmetric, although the Casimir of $SO(6)$ does commute with
$H_{eff}$ and can be used to classify states. This is why the would-be Goldstone modes are massive, but their masses are
protected.

It is also interesting that the result is independent of the size of the brane, since at large radius one would expect a blue-shift
which would increase the total energy of these states. However, these states are massless on the D-brane, and the blue-shift is offset by the fact that that the particles move on a bigger sphere, so that their energy is reduced by the increase in size of
the brane. If the particles were massive, this would not happen, because we would still have the rest mass of the particle.

Note also that the large giant gravitons expanded into AdS have a worldvolumes that lie ``close'' to the AdS boundary.  This is why local physics on such branes will be essentially described by local effective dynamics  in the boundary theory also. For some calculations we can even ignore the curvature of the $S^3$, if we are considering very massive states on the D-brane worldvolume, like massive strings.

\paragraph{Massive strings on $AdS_5$ branes: }  There is an attractive interpretation of the massive off-diagonal modes with masses (\ref{massivemass}) in terms of massive excitations of the strings on an $\ads{5}$.  Consider a single brane state (\ref{semi1}) which breaks the Yang-Mills gauge group to $SU(N-1) \times U(1)$.   Here the $SU(N-1)$ dynamics in the gauge theory is interepreted as reproducing the gravitational physics of the semiclassical AdS space in which a single probe brane described by the $U(1)$ is embedded.  In order to construct gauge invariant  massive excitations of the $\phi^i$, one needs oscillators of the form 
\begin{equation}
\phi^i \sim \begin{pmatrix}0 & a^\alpha & \cdots \\
b_\alpha&0 & \cdots \\
\vdots & \vdots & \ddots
\end{pmatrix}
\end{equation}
so that gauge invariant states can be constructed as 
\begin{equation}
\sum b^\dagger_\alpha (a^\alpha)^\dagger |0>_{A}
\end{equation}
More generally one can build states including ``gluons" of the $SU(N-1)$ dynamics. 
This construction is interpreted as a  realization of Gauss's law on the spherical brane in AdS.  To see this, recall that since the brane is compact, the two oppositely charged ends of an open string must end on it.   Since the string is oriented, we can interpret the $a^{\alpha}$ as (say) the outgoing end of the open string and $b_\alpha$ as the incoming piece.   The off-diagonal oscillators in the Yang-Mills are charged  under the $U(1)$ which describes the probe brane and also under the $SU(N-1)$ which realizes the AdS dynamics.  Pictorially the off-diagonal excitations leave the brane (the top left eigenvalue), explore the bulk AdS (the $SU(N-1)$ part of the $\phi^i$ matrix) and then return to the brane.   This is similar to the description of strings on branes in the Poincare patch of AdS space that are described in terms of the Coulomb branch of a dual theory on a plane.

 In the Yang-Mills theory, eq.~(\ref{massivemass}) indicates that such excitations will have a mass of order $\sim \sqrt{\lambda j}$ at large $\lambda$.    When $\lambda \sim g^2N$ is large, the dual spacetime has a very low curvature and thus the strings on a large spherical D-brane are essentially moving in flat space.  The proper energy of such a string (translated CFT units by multiplying by the AdS scale) will go as $E_p \sim  {\root 4 \of \lambda}$.  To relate this to the dual CFT we must multiply by the redshift factor arising from the relation between coordinate and global time. The metric in global coordinates of $AdS_5$ is given by
\begin{equation}
ds^2 = -(1+r^2)dt^2 + \frac1{1+r^2}dr^2 +r^2 d\Omega_3^2
\end{equation}
The proper energy at position $r$, $E_p$ is related to energy measured  as $E \sim \sqrt{1+r^2} E_p$.  Recall that the string is question is attached to brane with angular $J = Nj$, and such branes have a radius $r^2 = j$ \cite{goliath,akietal}.  Thus the energy of the string in CFT units is predicted by supergravity analysis to be $E \sim \sqrt{j}  {\root 4 \of \lambda}$.   This disagrees with the mass of $\sqrt{j \lambda}$ (\ref{massivemass}) predicted in the free field analysis on the Yang-Mills theory although the $\sqrt{j}$ dependence agrees.  Notice that agreement would be obtained if a full quantum treatment in Yang-Mills theory simply led to the replacement $\lambda \to \sqrt{\lambda}$.  This has been seen before in AdS/CFT Wilson loop calculations \cite{wilson}, and the similarity is not accidental since those computations were also dealing with the physics of the Coulomb branch, albeit on a plane.

\section{``Baryons" as D-branes}
\label{baryonsasbranes}

In the previous section we argued that many properties of spherical branes in $\ads{5}$ and the open strings on them are simply understood in terms of the analog of a Coulomb branch on a sphere.  In this section we turn to a more challenging problem -- the description of spherical branes that have expanded in the $S^5$ (giant gravitons).   It was proposed in \cite{BBNS} that such states are described in the dual field theory by baryon-like operators -- determinants and subdeterminants of scalar fields.  To be specific let $X,Y,Z$ be three complex scalar fields constructed from the six real chiral scalars of the Yang-Mills theory $\phi^i$ as $X=\frac{\phi^1+i\phi^2}{\sqrt{2}}$, $Y=\frac{\phi^3+i\phi^4}{\sqrt{2}}$ and
$Z=\frac{\phi^5+i\phi^6}{\sqrt{2}}$.  The giant gravitons are described by operators  
\begin{equation}
\CO_M  = \epsilon^{j_1 \cdots j_{M} i_{M+1}\cdots i _N}_{i_1\cdots i_{M} i_{M+1}\cdots i _N} Z^{i_1}_{j_1} \cdots Z^{i_{M}}_{j_{M}}
\Label{baryons}
\end{equation}
Here $M=N$ for a maximal size giant graviton. The group theoretic arguments of \cite{CJR} and the matrix model analysis of \cite{Btoy} confirm this.   

It has also been shown that the Yang-Mills theory contains excitations with the correct charges and conformal dimensions to describe strings propagating on the giant gravitons in spacetime \cite{BHLN}.  For example, strings propagating on a maximal giant graviton are proposed to be described by the operators like \cite{BHLN}:
\begin{eqnarray}
\CO^{Z,Y,X}_k &=& \epsilon_{i_1\cdots i_N}^{j_1\cdots
j_N}Z^{i_1}_{j_1}\cdots Z^{i_{N-1}}_{j_{N-1}} (Y^kXY^{J-k} 
)^{i_N}_{j_N}
\Label{opNeumann} \\
\CO^{Z,Y,Z}_k &=& \epsilon_{i_1\cdots i_N}  ^{j_1\cdots
j_N}Z^{i_1}_{j_1}\cdots Z^{i_{N-1}}_{j_{N-1}} (Y^kZY^{J-k}
)^{i_N}_{j_N}
\Label{opDirichlet}
\end{eqnarray}
Here the string of $Y$s builds up the worldvolume of a string excitation with momentum $J$ in a manner similar to the description of strings on pp-waves \cite{BMN} and on dibaryon states
\cite{BHK}.  The additional insertions of the ``impurities'' $X$ in (\ref{opNeumann}) and $Z$ in (\ref{opDirichlet}) correspond further oscillator excitations of the string.     As described in the dictionary given in \cite{BHLN}, (\ref{opNeumann}) should correspond to a fluctuation along the brane which therefore has Neumann boundary conditions, while (\ref{opDirichlet}) represents a transverse fluctuation that should have Dirichlet boundary conditions.  The missing piece of logic in demonstrating that ``baryons'' such as (\ref{baryons}) are actually D-branes is to show these Neumann and Dirichlet boundary conditions emerge dynamically in the Yang-Mills theory.
We will show this in the plane wave limit, where it is easy to compare with results for open string spectra on D-branes \cite{parklee,DP,kostasmarika,openstrings}

To do this, we must compute the normalized matrix elements
\begin{equation} \label{matrix}
M_{ij}=\frac{\bra \CO^*_i(x)\CO_j(0)\ket_{free+interacting}}
{\sqrt{\bra\CO^*_i(x)\CO_i(0)\ket_{free}} \sqrt{\bra\CO^*_j(x)\CO_j(0)\ket_{free}}
}
\end{equation}
where $i,j = 1 \cdots J$ represent the position of the impurity within the string.   The relevant part of  the $\CN$ SYM action  is
\begin{equation}
S={1 \over 2\pi g_s} \int d^4 x ~\tr\Bigl({1 \over 2}F_{\mu
\nu}F^{\mu \nu}+D_\mu Z D^\mu \overline{Z} +D_\mu Y D^\mu
\overline{Y}+D_\mu X D^\mu\overline{X} +V_D+V_F\Bigr)
\end{equation}
Where the D-term potential and the F-term potentials are
\begin{equation}
V_D=\frac{1}{2}\tr|[X,\overline{X}]+[Y,\overline{Y}]+[Z,\overline{Z}]|^2
\end{equation}
\begin{equation}
V_F=2\tr(|[X,Y]|^2+|[X,Z]|^2+|[Y,Z]|^2)
\end{equation}
According to the argument in Appendix B in \cite{Constable}, The
D-term and gluon exchange cancel at one loop order (this is based
on techniques in previous papers \cite{DHoker, Skiba} ), so to this order we
only need to consider the contributions from F-term.  The scalar
propagators are
\begin{equation}
\bra Z_i^j(x) \overline{Z}_k^l(0) \ket = \bra Y_i^j(x)
\overline{Y}_k^l(0) \ket =  \bra X_i^j(x) \overline{X}_k^l(0) \ket
= \delta^l_i \delta^j_k {2 \pi g_s \over 4 \pi^2}{1 \over |x|^2},
\end{equation}

\subsection{Neumann}
The matrix (\ref{matrix}) $M_{ij}$ consists of a free part, which is easy to compute, and an
interacting part.  The norms of the operators (\ref{opNeumann}) are
\begin{equation}
 \bra (\CO^{Z,Y,X}_k(x))^* \CO^{Z,Y,X}_k (0) \ket_{free}  = {(N-1)! ~ N!^2}{1 \over |x|^{2(N+J) }}N^{J}
 (\frac{2\pi g_s}{4\pi^2})^{N+J}  \equiv
{C \over |x|^{2 \Delta}},
 \end{equation}
where $C\equiv {(N-1)! ~ N!^2}N^{J}(\frac{2\pi g_s}{4\pi^2})^{N+J}
$ and $\Delta=(N+J) $. The factor of $(N-1)!$ counts the number of
contractions between the Zs.   $N!^2$ is the result of contracting
four $\epsilon$ tensors in pairs and $N^{J}$ results from $J$
loops from contracting the Ys and X planarly. It is obvious that the free off-diagonal part will be suppressed by powers of $N$ since we can not contract planarly, so the free
part in $M_{ij}$ is just identity matrix.
\begin{equation}
M_{ij}^{free}=\delta_{ij}
\end{equation}

To calculate the one-loop interacting piece we insert the
operator $-\frac{1}{2\pi g_s}V_F$ in the correlator, and then integrate the three point
function over the position of the inserted operator. The leading
large N contributions come from
$-2tr([X,Y][\overline{X},\overline{Y}])$ in the F-term.\footnote{There is also a non-vanishing constant contribution from
interaction of $Z's$ and $Y's$ as shown in appendix in
\cite{BHLN}, but this contribution is of order $g_s$, so is much
smaller in planar limit than the contribution of order
$g_s\frac{N}{J^2}$ we consider here. This small contribution is
argued to be related to one-loop open strings effect \cite{BHLN},
which we will not discuss here.} (The calculation is very similar
to those made with BMN operators in pp-wave settings \cite{BMN}). 
First consider the diagonal part of $M_{ij}$. For $1\leq k\leq J-1$, following the calculations in the appendix of \cite{BHLN} we find
\begin{eqnarray}
&& \bra \CO^{Z,Y,X}_k(x)^* \CO^{Z,Y,X}_k(0) \ket_{one-loop}   \\
 &=& \int d^4y \bra  \CO^{Z,Y,X}_k(x)^* \CO^{Z,Y,X}_k(0)
= 
\Bigl(- {2Ng_s \over{ \pi }}\Bigr){C \over
|x|^{2\Delta}}\log(|x|\Lambda) ~~~~~~~~~~~~~~~~~~~~~~~~~~~~ \nonumber
\end{eqnarray}
For $k=0$ or $k=J$ only $\tr(XY\overline{Y}\overline{X})$ or
$\tr(YX\overline{X}\overline{Y})$ contribute, so we have one half
of the result
\begin{eqnarray}
&&\bra \CO^{Z,Y,X}_0(x)^* \CO^{Z,Y,X}_0(0) \ket_{one-loop}=\bra
\CO^{Z,Y,X}_J(x)^* \CO^{Z,Y,X}_J(0) \ket_{one-loop}
 \nonumber \\
&=& \Bigl(- {Ng_s \over{ \pi }}\Bigr){C \over
|x|^{2\Delta}}\log(|x|\Lambda)
\end{eqnarray}
Now we compute off-diagonal  piece. The calculation is similar to
above except $\tr(XY\overline{X}\overline{Y})$ or
$\tr(YX\overline{Y}\overline{X})$ contribute, so we have opposite
sign contribution. We find (for $1\leq k\leq J$)
\begin{equation}
\bra \CO^{Z,Y,X}_{k-1}(x)^* \CO^{Z,Y,X}_k(0) \ket_{1-loop}  \bra \CO^{Z,Y,X}_k(x)^* \CO^{Z,Y,X}_{k-1}(0)
\ket_{1-loop} 
= \Bigl( {Ng_s \over{ \pi }}\Bigr){C \over
|x|^{2\Delta}}\log(|x|\Lambda)
\end{equation}
In summary, the  matrix of overlaps (\ref{matrix}) at 1-loop is
\begin{equation}
M_{one-loop} = - {Ng_s \over{ \pi }}\log(|x|\Lambda)
\left(
\begin{matrix}
1 & -1 & 0 & 0 & \cdots \cr -1 & 2 & -1 & 0 &
\cdots \cr 0 & -1 & 2 & -1 & \cdots \cr \vdots& \ddots & \ddots &
\ddots & \vdots \cr 0 & \cdots & 0 & -1 & 1 \cr 
\end{matrix}
 \right)
\end{equation}
By diagonalizing this one-loop planar mixing matrix we can obtain a basis of energy  eigenstates. As argued in \cite{openstrings} the eigenstates with eigenvalue $n$ are superpositions of the operators (\ref{opNeumann}) with coefficients $\cos(\pi n k/J)$.     The cosine factor directly describes a fluctuation of the string with Neumann boundary conditions on the brane.  The 1-loop anomalous dimensions of the orthogonal operators are extracted from the diagonalized overlap matrix by removing the $\log$ and diving by $2$.

\subsection{Dirichlet}
We can repeat the analysis for the operators (\ref{opDirichlet}). Everything goes through as above, if the insertion $Z$ is not at the edges.     However, if $Z$ is at the edge of the string of $Y$s, say at $k=0$ we can immediately see that the operator changes character \cite{Bgiant}. It factorizes as
\begin{equation}
\CO^{Z,Y,Z}_0= \epsilon_{i_1\cdots i_N} ^{j_1\cdots
j_N}Z^{i_1}_{j_1}\cdots Z^{i_{N-1}}_{j_{N-1}} (ZY^{J}
)^{i_N}_{j_N}=\frac{1}{N} (N!\det(Z))\tr(Y^J)
\end{equation}
where $N!\det(Z)=\epsilon_{i_1\cdots i_N} ^{j_1\cdots
j_N}Z^{i_1}_{j_1}\cdots Z^{i_N}_{j_N}$.  We interpret this as saying that state has factorized in a giant graviton ($\det Z$) with a nearby closed string ($\tr(Y^J)$).  The norm is
\begin{equation}
 \bra (\CO^{Z,Y,Z}_0(x))^* \CO^{Z,Y,Z}_0 (0) \ket_{free}  = J {N!^3}N^{J-2}
 (\frac{2\pi g_s}{4\pi^2})^{N+J}{1 \over |x|^{2(N+J) }}
 \end{equation}
The free off-diagonal part is also suppressed by powers of $N$ and
the free part matrix $M_{ij}$ is identity matrix
$M_{ij}^{free}=\delta_{ij}$ as in the case of Neumann boundary
condition.

Now consider one-loop part of the matrix element involving the $k=0,J$ impurities.
\begin{eqnarray}
\bra \CO^{Z,Y,Z}_0(x)^* \CO^{Z,Y,Z}_0(0) \ket_{one-loop} = \int
d^4y \bra \CO^{Z,Y,Z}_0(x)^* \CO^{Z,Y,Z}_0(0) (\frac{1}{\pi
g_s})\tr([Y,Z][\overline{Y},\overline{Z}])(y) \ket
\end{eqnarray}
The leading large $N$ contribution turns out to be
\begin{eqnarray}
\bra \CO^{Z,Y,Z}_0(x)^* \CO^{Z,Y,Z}_0(0) \ket_{1-loop}
&\sim& (N-1)!^3\bra \tr(Z\overline{Z})
\tr(Y^J)\tr(\overline{Y}^J)\tr([Y,Z][\overline{Y},\overline{Z}])\ket
\nonumber \\ 
&\sim & (N-1)!^3 J^2N^{J+1}
\end{eqnarray}
So the ratio between the free and interacting parts
\begin{eqnarray}
\frac{\bra \CO^{Z,Y,Z}_0(x)^* \CO^{Z,Y,Z}_0(0)
\ket_{one-loop}}{\bra (\CO^{Z,Y,Z}_0(x))^* \CO^{Z,Y,Z}_0 (0)
\ket_{free}} \sim J \frac{g_s}{\pi}\log(|x|\Lambda)
\end{eqnarray}
This is suppressed compared to the matrix elements of the overlap matrix $M$  for impurities in the interior which are all of order $N$.

Finally, we compute the off-diagonal elements of $M$ connecting the $k=0,J$ impurities at the edge of the string and  impurities in the interior $\bra \CO^{Z,Y,Z}_0(x)^*
\CO^{Z,Y,Z}_1(0) \ket_{1-loop}$. (The overlap with $k$ deeper in the interior is further suppressed.)
The leading large $N$ power is
\begin{eqnarray}
 \bra \CO^{Z,Y,Z}_0(x)^* \CO^{Z,Y,Z}_1(0) \ket_{1-loop}
&\sim& (N-1)!^3 \bra \tr(\overline{Y}^J)
\tr(\overline{Z} YZY^{J-1}) \tr([Y,Z][\overline{Y},\overline{Z}])
\ket \nonumber \\
&\sim & J (N-1)!^3 N^{J+2}
\end{eqnarray}
So in the leading large $N$ order
\begin{eqnarray}
\frac{\bra \CO^{Z,Y,Z}_0(x)^* \CO^{Z,Y,Z}_1(0)
\ket_{1-loop}}{
\sqrt{\bra \CO^{Z,Y,Z}_0(x)^* \CO^{Z,Y,Z}_0(0)
\ket_{free}}
\sqrt{\bra \CO^{Z,Y,Z}_1(x)^* \CO^{Z,Y,Z}_1(0)
\ket_{free}}
}
\sim
\frac{g_s}{\pi}\sqrt{NJ}\log(|x|\Lambda)
\end{eqnarray}
In summary the one-loop overlap matrix (\ref{matrix}) is
\begin{equation}
M_{1-loop} = - {Ng_s \over{ \pi }}\log(|x|\Lambda)
\left(
\begin{matrix}
 O(\frac{J}{N}) & O(\sqrt{\frac{J}{N}}) & 0 & 0 &
\cdots \cr O(\sqrt{\frac{J}{N}}) & 2 & -1 & 0 & \cdots \cr 0 & -1
& 2 & -1 & \cdots \cr \vdots& \ddots & \ddots & \ddots & \vdots
\cr 0 & \cdots & 0 & O(\sqrt{\frac{J}{N}}) & O(\frac{J}{N})\cr 
\end{matrix}
\right)
\end{equation}
At large N the contributions from the edges of the string $k=0,J$ decouple which is in agreement with the observation at the beginning of this section that these operators describe a state with a giant graviton and separate closed string.  It was argued in \cite{openstrings} that a matrix of overlaps of the form of the remainder of $M$ after eliminating $k=0,J$ diagonalizes to describe string excitations with Dirichlet boundary conditions on a brane, namely excitations where the endpoints do not move.

\subsection{Open strings on smaller D-branes}

Above we were discussing the leading large $N$ contributions in the Yang-Mills to the anomalous dimension of operators dual to excitations of strings attached to a maximal size D-brane.  As mentioned we neglected an $O(1)$ 1-loop piece which is, however, the dominant contribution to the anomalous dimension of the ground state of the string.  In \cite{BHLN} this contribution was shown to be
\begin{equation}
\frac{(J-1)g_s}{\pi} \, .
\Label{maxzerodim}
\end{equation}
This contribution to the  dimension should arise in spacetime from  a one-loop open string effect since it has one less power of $N$ than the leading terms in the dimension of excited strings \cite{BHLN}.   Here we study the 1-loop contribution to the dimension of operators describing the ground of a string on a giant graviton of less than maximal size, namely $\CO_M$ in (\ref{baryons}) with $M<N$.   Quite surprisingly, we find the one-loop anomalous
dimension of the ground state of the string grows like $\frac{(N-M)g_s}{\pi}$.  The spacetime interpretation of this is not entirely clear, but the result will be of use in later sections.

The dual operator of an open string attached to a non-maximal giant
graviton is \cite{BBNS,CJR,BHLN}
\begin{equation}
\CO\equiv\epsilon_{i_1\cdots i_M }^{j_1\cdots j_M
}Z^{i_1}_{j_1}\cdots Z^{i_{M-1}}_{j_{M-1}}(Y^J)^{i_M}_{j_M}
\end{equation}
We compute the free and one-loop two point function $\langle \CO
\overline{\CO}\rangle$ in  the leading order at large $N$ and $M$,
but we do not assume $N-M$ to be large.\footnote{Here we assume $J>1$.  When $J=1$ additional terms contribute, and the one-loop
anomalous dimension cancels and does not grow. This is just as
expected since when $J=1$ this operator is a BPS operator so there
should be no anomalous dimension.}  After calculations
using formulae in appendix \ref{property}, we find the free and
one-loop two point function.
\begin{eqnarray} 
\label{sub:free}
\langle \CO(x) \overline{\CO} (0)\rangle_{free} 
&=& 
N^{J+1}
\frac{(N-2)!(M-1)!^2(M-1)}{(N-M)!}(\frac{g_s}{2\pi|x|^2})^{M+J-1}
 = \frac{C}{|x|^{2\Delta}}
 \nonumber  \\
 \label{sub:one-loop}
\langle \CO(x) \overline{\CO} (0)
\rangle_{1-loop}&=&-\frac{2g_s}{\pi}(N-M+J-1)
\frac{C}{|x|^{2\Delta}}\log(|x|\Lambda)
\end{eqnarray}
Here $\Delta=M+J-1$ is the free field conformal dimension. From
(\ref{sub:free}) and (\ref{sub:one-loop})  the
one-loop anomalous dimension is
\begin{equation} 
\delta\Delta_{one-loop}=\frac{(N-M+J-1)g_s}{\pi} \, .
\Label{sub}
\end{equation}
So when $M=N$ so that the brane is of maximal size we reproduce the result $\frac{(J-1)g_s}{\pi}$ in
\cite{BHLN}.

\section{Multi-brane states and Gauss' Law}

The arguments in \cite{BHLN} and in Sec.~4 together establish that determinants and sub-determinants are D-brane operators of Yang-Mills theory, supporting open string excitations at large N.      A system of $G$ such D-branes should dynamically realize a new gauge theory associated with the low-energy excitations of these strings.  This local gauge theory will be local on an emergent space, the D-brane worldvolume, which is not embedded in the space on which the original Yang-Mills theory is defined.   To study this we must first propose a description of a multi-D-brane state.  An approximately orthogonal basis of low conformal dimension states (corresponding to supergravity particles) can be constructed from products of single-trace operators, with number of traces being an approximate quantum number counting the number of particles \cite{jthroat}.  Thus, a natural guess for an operator describing a system of Yang-Mills D-branes is a product of sub-determinants, with the open strings stretching between the branes represented as defects with indices entangled between the D-brane operators.   Since the D-branes are compact, Gauss' law should be realized by the restriction that as many strings should start as end on any given brane in order that the net charge on the brane be zero.

It will turn out that for a state with two branes, one of them of maximal size, the naive product operator is the correct one.   It is instructive to first see how Gauss' Law is realized in this special case.   Naively one would expect four states -- one string on each brane and two (oriented) strings running between them.  However,  Gauss' Law forbids the strings between the branes.

As in the previous section, a string attached to a single brane is described by \cite{BHLN}
\begin{equation}
{\cal O}\equiv\epsilon^{j_1j_2\cdots j_M}_{i_1i_2\cdots
i_M}Z^{i_1}_{j_1}Z^{i_2}_{j_2}\cdots
Z^{i_{M-1}}_{j_{M-1}}(Y^J)^{i_M}_{j_M} \, .
\end{equation}
The natural guess for the operator describing a maximal brane and a smaller brane with a string attached to it is:
\begin{equation}
\CO_{22}\equiv\epsilon_{i_1\cdots i_N }^{j_1\cdots j_N
}Z^{i_1}_{j_1}\cdots Z^{i_{N}}_{j_{N}}\epsilon^{l_1\cdots l_M
}_{k_1\cdots k_M }Z^{k_1}_{l_1}\cdots
Z^{k_{M-1}}_{l_{M-1}}(Y^J)^{k_M}_{l_M} \, .
\end{equation}
As pointed out in \cite{BHLN}, we can construct another operator
from the above by exchanging the upper (or lower) index
of $Y^J$ with one of the $Z$'s in the other brane. This gives
\begin{equation}
\CO_{12}\equiv\epsilon_{i_1\cdots i_N }^{j_1\cdots j_N
}Z^{i_1}_{j_1}\cdots Z^{k_{M}}_{j_{N}}\epsilon^{l_1\cdots l_M
}_{k_1\cdots k_M }Z^{k_1}_{l_1}\cdots
Z^{k_{M-1}}_{l_{M-1}}(Y^J)^{i_N}_{l_M} \, .
\end{equation}
The operator thus constructed looks like a single open string
stretching between two branes.  This would violate Gauss' Law.  

Now observe that in $\CO_{12}$ if $k_M\neq i_N$,
then there is an $x<N$ such that $k_M= i_x$. There will be two
$Z$'s in the operator with the indices $Z^{i_x}_{j_x}$ and
$Z^{i_x}_{j_N}$. These terms will then be cancelled by the
antisymmetrization in the summation of $j$'s. We are left with
terms that have $k_M=i_N$. It is then easy to see that these two
operators are related by
\begin{equation}
{\cal O}_{12}=\frac{1}{N}{\cal O}_{22}.
\end{equation}
This is because in the operator $\CO_{22}$ there must be a $1\leq
x\leq N$ that $i_x=k_M$. We have $N$ choices of $x$ and each of
them gives us the result $\CO_{12}$ according the above argument.  So we see 
that a combinatorial identity causes a state that would apparently violate Gauss' Law
to be equivalent to a linear combination of states that do not do so.   

We would like to go beyond this simple special case to a general analysis that applies to arbitrary systems of branes and open strings.   In general, it turns out that products of antisymmetric representations are not a useful orthogonal basis -- indeed, they decompose into sums of irreducible representations that have non-zero overlaps -- and they do not obviously realize Gauss'  Law.    Below, following ideas in \cite{CJR,Btoy}, we describe an alternate basis of operators for branes with open string excitations and show that Gauss' law is realized almost automatically.

\subsection{Multi-brane states}

It is helpful to start with the free fermion description of half-BPS states.  The fermions wave functions are identified with characters of different irreps of $U(N)$.\footnote{Of course we should really be using $SU(N)$ in order to apply the AdS/CFT correspondence.   However, the differences between the two analyses are suppressed by powers of $N$.  Also, one can do the calculations in $U(N)$ and systematically eliminate the $U(1)$ diagonal degrees of freedom, since they are decoupled.  Thus, in the remainder of this paper we will neglect this issue which been explored in \cite{DeM}.}   Recall from Sec.~2 that $k$ coincident D-branes expanding onto $S^5$ are described a fermi sea with $k$ holes, and that this is a configuration corresponding to a $k$ column Young tableau.  We can think of these as tensors in the variable $Z^i_j$, where the Young tableau realizes the symmetry under permutations of the upper matrix indices of $Z$. Since the $Z$ are indistinguishable bosons, the upper and lower indices have  to transform identically under permutations (see, e.g.,  \cite{BQHE}).  In other words
$
Z^{i_1}_{j_1}Z^{i_2}_{j_2}
$
is invariant under the double exchange of sub-indices on the labels $1\leftrightarrow 2$.

When the upper and lower Young tableaux are the same (as they are for a state created from a product of $Z$z only), a $U(N)$ singlet is formed by taking a trace, i.e., contracting all upper indices with all lower indices.  For more general states that are not half BPS, we must add more matrix degrees of freedom in addition to the $Z$ field.   In particular, we will think of open string states as words constructed from the alphabet of field variables and then treat them as ``composite'' matrices  with one upper and one lower index per string $(M_\alpha)^i_j$.  This leads to a partonic description of the string in the free field limit.  To build states of strings on D-branes, we separately tensor the upper and lower indices of the D-brane operator with the (multiple) string states, and then decompose the result in representations of $U(N)$.   Roughly, this means that we start with the Young tableau for the D-branes and add to it an additional box for each string state we are considering.  In this picture the shape of the tableau we start with defines the system of branes and adding strings changes the shape of the tableau.  (The next section will take a slightly different approach which is essentially equivalent when the branes are large.)

\begin{figure}[ht]
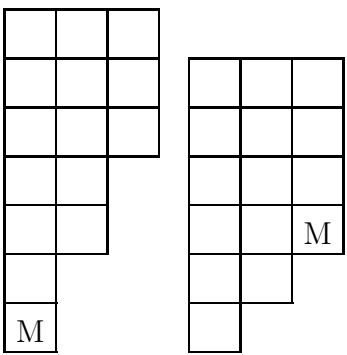

\begin{equation*}
\begin{Young}
& & \cr
& & \cr
& & \cr
&\cr
&\cr
\cr
M\cr
\end{Young} \quad\begin{Young}& & \cr
& & \cr
& & \cr
& & M \cr
&\cr
\cr
 \end{Young}
\end{equation*}
\caption{One string beginning in one giant graviton and ending in another one}\label{fig:double}
\end{figure}

\begin{figure}[ht]
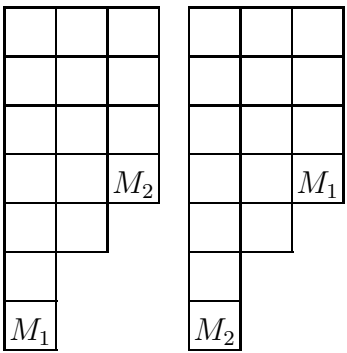

\begin{equation*}
\begin{Young}
& &\cr
& & \cr
& & \cr
& & $M_2$\cr
&\cr
\cr
$M_1$ \cr
\end{Young} \quad\begin{Young}
& & \cr
& & \cr
& & \cr
& & $M_1$\cr
& \cr
\cr
$M_2$\cr
 \end{Young}
\end{equation*}
\caption{Two strings stretching between two giants with opposite orientations}\label{fig:glaw}
\end{figure}

Since the composite string degree of freedom $M$ can be distinguished from $Z$, the upper and lower indices of $M$ need not be in the same box of a Young tableau --  the trace can be non-vanishing even if the indices are ``entangled'' in this way.    Thus, for each state, we can consider a double Young tableaux representation, where we keep track of both upper and lower indices
Pictorially, the string $M$ begins (ends) on the brane represented by the Young tableau column which contains the box associated to the upper (lower) index of M.   For example, see Fig.~\ref{fig:double}.
In this case, the upper and lower indices correspond to different Young 
tableaux. If one tries to make a gauge invariant state by contracting upper and lower indices
one gets zero, since there is no $U(N)$ singlet associated to the tensor product of these two irreducible representations.  This is expected by Gauss' Law, which forbids a single string stretched between two distinguishable compact branes.   However, we can consider adding more strings as in Fig.~\ref{fig:glaw}. We see in this case that the tableaux of upper and lower indices have the same shape but the boxes corresponding to upper and lower indices of an individual string might be different.  So now when we contract the indices there is a singlet, which is consistent with Gauss' Law: a pair of oppositely oriented strings can run between two compact branes.

In this construction, the requirement that operators be $U(N)$ invariant forces appropraite contractions of upper and lower indices to obtain singlet states.  This forces the upper and lower Young tableaux to have the same shape, or equivalently, to have the same number of strings starting and ending on each brane.  This is Gauss' law.  The argument can be extended to multiple D-branes expanding into $AdS_5$ as well. There, strings beginning and ending on one of those giants will add boxes to Young tableaux rows that represent D-branes. This  symmetry between the exchange of columns and rows
in the description of states amounts to particle/hole duality and suggests that whatever happens for the AdS giants also happens for the $S^5$ giants.

At this point, it would be convenient if we could perform all the calculations with the given states above.
However, we have not found a way to do that systematically.  There is another basis of states which seems to be equivalent to the one described above, and we will work with it in the following sections because the combinatorics is easier to handle. We give some arguments suggesting the equality of the two prescriptions.  It is clear that the basis of operators described above seems to have the right intuition regarding strings stretching between different D-branes, so we will assert at this point without proof that it is the right solution to the problem of classifying all strings stretching between D-branes and counting them.

\subsection{Strings stretching between D-branes}

So far, we have seen how to build the states that are associated to 
different strings stretching between D-branes from a ``pictorial'' point of view. Now we want to be systematic and have not found a way to do it in the basis mentioned before.  For this purpose, as explained above, we will use a slightly different formalism from the previous section: (a) we will fix the tableau shape and replace some boxes (in arbitrary places on the Young diagram) with open strings (rather than adding open string boxes, thus changing the tableau shape), and (b) we will only have a tableau permuting the lower $U(N)$ indices (this is essentially considering the relative permutation between upper and lower indices in the methods of the previous section).   The results of the two formalisms hopefully will be identical and cover all possible operators near the giant graviton configuration systematically when the number of boxes is of order $N$.

The basic tool that we will use is the isomorphism between gauge invariant operators of unitary groups  and representations of the permutation group $S_n$.  (See \cite{groupbook}, \cite{CJR} and references therein.)

Recall that every irreducible representation of $S_n$ is associated to a Young diagram with $n$ boxes.  A tableau is constructed by filling the diagram with numbers $1 \cdots n$.  Let $H \in S_n$  be the {\it horizontal subgroup} of permutations of $1 \cdots n$ that preserves the rows of a Young diagram, and let $V \in S_n$ be the column-preserving {\it vertical subgroup}.   A  {\it Young element} of the group algebra of $S_n$ is associated to a Young tableau as
\bea
c_{{\rm Young}} = \sum_{p\in H} \sum_{q\in V} (-)^{\sign(q)} p\cdot q
\Label{cYoung}
\eea
where $\sign(q)$ is $\pm 1$ for even/odd permutations.    For example, the Young element of Fig. a
is given by
\bean
[I + P_{12}] \cdot [I - P_{13}] = I + P_{12} - P_{13} - P_{12} P_{13}
\eean
where $P_{ab}$ exchanges the labels $a$ and $b$.   The $n!$ Young elements associated to a given Young diagram are linearly dependent and span a sub-algebra of the complete group algebra.   A complete basis for the group algebra is obtained by taking the union of the sub-algebra bases obtained for each irreducible representaion of $S_n$.

Now as in \cite{CJR}, given a Young tableau of $S_n$ and related Young
elements  we can contruct a gauge invariant operator from
$n$ unitary matrices $\Phi_1,...,\Phi_n$ as
\bea
{\cal O}_{Young}=\sum_{i_1,...,i_n=1}^N \sum_{p\in H} \sum_{q\in V} (-)^{\sign(q)}
(\Phi_1)^{i_1}_{i_{p\cdot q(1)}}(\Phi_2)^{i_2}_{i_{p\cdot q(2)}} ...
(\Phi_n)^{i_n}_{i_{p\cdot q(n)}}~.
\Label{Young-op}
\eea
For example, the Young tableau (a) in  Fig.\ref{f:3box}  will
give
\bean
{\cal O}= \Tr \Phi_1 \Tr \Phi_2\Tr \Phi_3- \Tr \Phi_1 \Phi_3\Tr \Phi_2 
+ \Tr \Phi_1 \Phi_2 \Tr \Phi_3- \Tr \Phi_1 \Phi_3 \Phi_2
\eean
If we set all $\Phi_i=\Phi$, a single complex scalar of the $\CN$ Yang-Mills theory, ${\cal O}_{Young}$ in (\ref{Young-op}) will be  a ${1\over 2}$-BPS operator as proposed in \cite{CJR}.\footnote{Of course if $\Phi_i \in SU(N)$, traces  of a single matrix will vanish.}

Following \cite{BBNS,CJR,Bgiant} we will interpret such 1/2-BPS tableaux that are largely antisymmetric, with columns of length of order the rank of the $SU(N)$ gauge group as collections of spherical D-branes that have expanded into the $S^5$ factor of $\ads{5} \times S^5$.   Columns of length larger than $N$ cannot exist because of antisymmetry between the rows.      One very special and simple case is the  product of a single-column tableau of length $N$ and any other single column tableau -- this is simply a joined 2-column tableau by the standard rules of multiplication of representations of the permutation group.  In this simple case, the operators we construct will be equivalent to the naive product of operators corresponding to a maximal size D-brane (a determinant) and any other D-brane (sub-determinant).
Extending this interpretation, we propose:
\begin{center}
\parbox[t]{5.5 in}{
{\sl Given a collection of ${1\over 2}$-BPS D-branes and the related Young diagram with $m$ boxes, a state with $k$ open strings is created by the Young operator ${\cal O}_{Young}$ associated to the tableau made by filling $k$ boxes in arbitrary locations with open string defects and $m-k$ boxes with a complex scalar $\Phi$.}} \\
\end{center}
Excitations of other members of a 1/2 BPS multiplet are constructed by acting on the Young tableau and operator with the broken SUSY generators. To test our proposal we will answer the following questions:
(1) How many independent $k$ string states can we construct?  Do these states and their number satisfy the restrictions imposed by Gauss' law on the D-branes?
(2) Are these operators approximately orthogonal at large $N$?

It is useful to start with a simple example.    Consider the Young diagrams in Fig.~\ref{f:3box} and construct operators by replacing $1,2 \to \Phi$ and $3 \to Y$.  (We only list three of the six Young tableaux and one of the three possible substitutions by $\Phi$ and $Y$ because the other choices give identical operators.) These states are not heavy enough to be excited D-branes, but they will be instructive nevertheless.  Alternatively we could imagine that both columns have $m \sim N$ boxes above them which have been suppressed for simplicity --  we could then think of this physically as two giant gravitons with an open string attached to them.   The three Young operators are 
\bean
Y_a & = & (\Tr \Phi)^2 \Tr Y- \Tr \Phi \Tr \Phi Y+ \Tr \Phi^2 \Tr Y
- \Tr \Phi^2 Y\\
Y_b & = & (\Tr \Phi)^2 \Tr Y- \Tr \Phi^2 Y\\
Y_c & = & (\Tr \Phi)^2 \Tr Y- \Tr \Phi^2 \Tr Y+
 \Tr \Phi \Tr \Phi Y- \Tr \Phi^2 Y
\eean
and there is one  relation among them
\bean 
2 Y_b= Y_a+ Y_c
\eean
Thus there are two independent states, which we can take to be created by the tableaux in Fig.~\ref{f:3box}a and \ref{f:3box}c.   This is consistent with Gauss' Law for compact D-branes.  A single string starting on a compact brane must also end on it so that the flux produced by the endpoint has both a  source and a sink.   By contrast there are four oriented string states on and between two noncompact D-branes.      A direct enumeration of Young operators and their relationships is intractable in the general case, so in what follows we will approach the problem of finding independent excited D-brane states more abstractly.

\begin{figure}

  \begin{center}
 \epsfysize=1.5in
   \mbox{\epsfbox{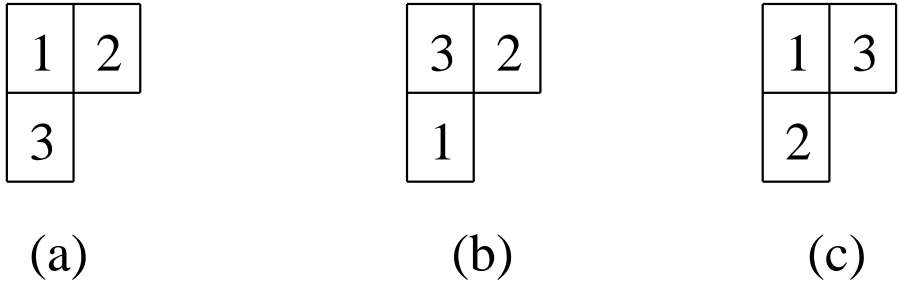}}
    \caption{The three possible Young tableau of $S_3$ that give different operators under the replacement $1,2 \to \Phi$ and $3 \to Y$.
 }\Label{f:3box}
 \end{center}
\end{figure}


Before ending the introduction of this section, we would like to give
a remark about the formula (\ref{Young-op}), where  
 upper indices are fixed and permutations act only on lower indices.
This approach used only one Young diagram of lower indices
and  is different from the two Young diagram method given in section 5.1.
However, it can be shown that these two methods are equivalent to 
each other. The reason is  the following. The sum 
$\sum_{s\in up} \sum_{t\in down} \Phi^{I_{s(i)}}_{I_{t(i)}}$
can be rewritten as 
$\sum_{s\in up} \sum_{t\in down} \Phi^{I_{i}}_{I_{ts^{-1}(i)}}$.
Summing over all $s,t$ is equivalent to acting with $c_{d} \cdot c_{u}$
on lower indices like (\ref{Young-op}) where $c_{u}$ and $c_d$ are 
corresponding Young elements (\ref{cYoung}). There are some basic
mathematical facts regarding  Young elements that can be found in \cite{groupbook}.
First, if the upper Young tableau  and lower Young tableau belong to different
Young diagrams then $c_{d} \cdot c_{u}=0$. This explains the intuitive
picture that both upper and lower index Young diagrams must have same
shape to get nonzero results. Second, for the upper and lower Young
tableaux of same Young diagram, the product can be expanded into
the linear combination of Young elements, i.e.,
\bean
c_{d}^{\a} \cdot c_{u}^{\a}=\sum_{i\in \a} f_i c_{i}^{\a}
\eean 
where we use explicitly $\a$ to denote a given Young diagram.
Since formula (\ref{Young-op}) has included all permutations
$c_{i}^{\a}$,  operators given by the
two Young diagram method of section 5.1 will just be linear combinations
of those with the one Young diagram method used in this section.

\subsubsection{A formulation of the problem of counting strings}

Consider again a single string attached to a system of two D-branes.  The Young diagram consists of two columns with $n = n_1 + n_2$ boxes and there are $n!$ ways to label the boxes with numbers $1 \cdots n$ to get a Young tableau.     To construct the relevant Young operators we can replace  $1,2,...,(n-1)\to X$ and $n\to Y$.   (Making any other choice of replacement of labels will lead to the same set of operators.)    The resulting set of operators will be invariant under permutations of the labels  $1 \cdots (n-1)$ by  $S_{n-1} \in S_n$.   Now recall that each assignment of labels $1 \cdots n$ to a tableau is associated to a specific Young element of the group algebra (\ref{cYoung}) which is not in general invariant by itself under $S_{n-1}$. Therefore  finding an independent basis of operators with two branes and one open string excitation is equivalent to finding a basis for the $S_{n-1}$ invariant subalgebra within the algebra generated by the Young elements $Y_i$  (i.e., $\sigma (\sum_i a_i Y_i) \sigma^{-1}= \sum_i a_i Y_i$ for $\forall \sigma \in S_{n-1}$).   This prescription is easily generalized.   For example, if the excitation involves two open string states $Y_1,Y_2$ with $Y_1\neq Y_2$ we must enumerate $S_n$ group algebra elements that are invariant under $S_{n-2}$, but if $Y_1= Y_2$, we seek group  algebra elements under $S_{n-2}\times S_2$. 

This splitting might seem artificial at first sight. However, when we compute correlators in two point functions, in the end we sum over all permutations of {\em identical} objects, which is just the statement that we compute these using  Wicks theorem. This will give us sums over subgroups of $S_n$ that permute the matrices which are to be treated as identical, but  will not sum over permutations that exchange different matrices: these do not appear in the free field contractions. 

\subsubsection{Counting the open string states}

We have reduced the problem of counting the number of $k$ string states on a collection of D-branes to enumerating elements of the $S_n$ group algebra that are invariant under subgroups of $S_n$.   Our task is aided by the following facts about the permutation group \cite{groupbook}:
\begin{itemize}

\item (1) A given Young diagram is associated to a single irreducible representation
of $S_n$ with dimension $d$ (which can be determined from the diagram) and unitary  representation matrix $D^R_{ij}(s)$ for $s\in S_n$.  
There are  {\sl orthogonality relationships} among these
unitary matrices: 
\bea 
\sum_{s\in G} D^{(\mu)}(s)_{il}  \O { D^{(\nu)}(s)_{sm}}
={N_G \over d_{\mu}} \delta_{is} \delta_{lm} \delta_{\mu\nu}  \, .
\Label{orth-D}
\eea
The  overline denotes complex conjugation, $N_G$ is the dimension of the group $G$ and $d_\mu$ is the dimension of the irreducible representation $\mu$.

\item (2) The $n!$ ways of labeling the $n$ boxes of a Young diagram to give a Young tableau are associated with $n!$ Young elements of the $S_n$ group as in (\ref{cYoung}).  If the diagram is related to a $S_n$ representation of dimensions $d$, there are $d^2$ linearly independent Young elements. 
A basis for them is:
\bea
A^R_{ij} =\sum_{s\in S_n} \O D^{R}_{ij}(s) \cdot s 
 \Label{A-def}
\eea
It is easily checked that 
\bea
g \cdot  A^R_{ij} \cdot g^{-1}=\O { D^{R}(g^{-1})}_{ik}  A^R_{kl}\O { D^{R}(g)}_{lj}= A^R_{kl}  D^{R}(g)_{ki}\O { D^{R}(g)}_{lj}\Label{action-1}
\eea

\item (3) Under the decomposition $S_{n_1+n_2}\to
S_{n_1}\times S_{n_2}$, the  
representation $R$ of $S_{n_1+n_2}$ decomposes as
\bea
R|_{S_{n_1+n_2}} =\sum_{R_1,R_2} C^{R}_{R_1,R_2} R_1|_{S_{n_1}}\times
R_2|_{S_{n_2}} \Label{Decom} \, 
\eea
where  the Littlewood-Richardon coefficients $C^{R}_{R_1,R_2}$ count the multiplicity of each product representation within $R$.  By  Frobenius Reciprocity,  $C^{R}_{R_1,R_2}$ equivalently counts the number of times  $R$ appears in the product of $R_1$ and $R_2$ (i.e., $R_1\times R_2\uparrow^{S_{n_1+n_2}}$)
 \cite{groupbook}.  The latter can be calculated easily by the Littlewood-Richardson rule for multiplying Young tableaux (see, e.g., Appendix A of \cite{groupbook2}).\footnote{\label{foot}The Littlewood-Richardson rule for multiplying Young diagrams $R_1 \otimes R_2$ is as follows.  Label each box of the ith row of $R_2$ by $i$.    Attach the boxes of of $R_2$ to $R_1$ in any way that gives a valid Young diagram, while obeying two constraints: (1) No boxes with the same number in them can appear in the same column, (2) Reading from right to left and starting with the top row and working down, the number of times the integer $i$ appears must be no less than the number of times the integer $i+1$ appears.\cite{groupbook2}   The Littlewood-Richardson coefficients count the number of times a particular Young diagram appears in this procedure.}  Examples of carrying out this procedure for tableaux with two columns are given in Fig.~4 ($S_n \to S_{n-1}$), Fig.~5 ($S_n \to S_{n-2}$) and Fig.~6 ($S_n \to S_{n-1} \times S_2$).
\end{itemize}

\begin{figure}
  \begin{center}
 \epsfysize=1.5in
   \mbox{\epsfbox{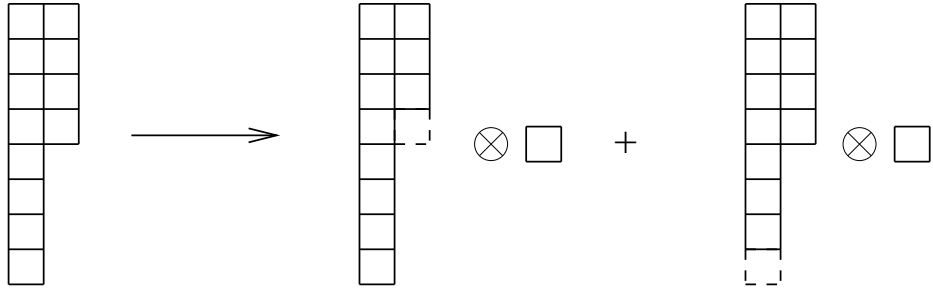}}
    \caption{The decompsition of $S_n$ to $S_{n-1} \times S_1$.  The dashed boxes indicate where the single box of $S_1$ can attach (see footnote \ref{foot}).}\label{f:2column-d}
 \end{center}
\end{figure}

\begin{figure}
  \begin{center}
 \epsfysize=1.5in
   \mbox{\epsfbox{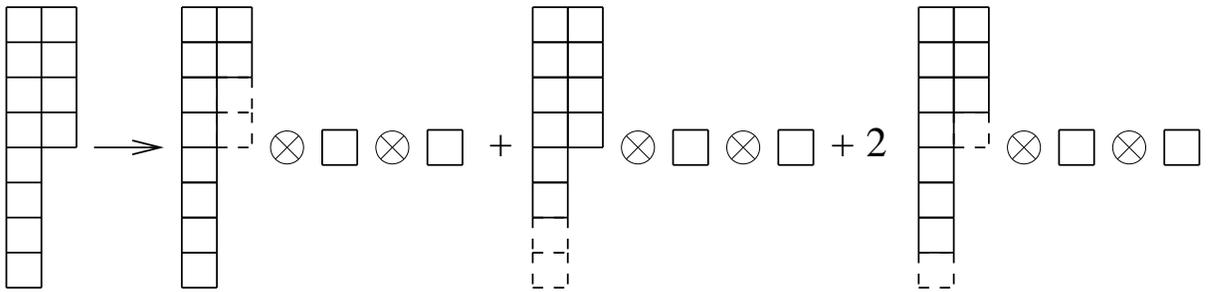}}
    \caption{The decomposion  of $S_n\to S_{n-2} \times S_1 \times S_1$.  The dashed boxes indicate where the single boxes of the two $S_1$s can attach.  The factor of $2$ indicates that the third $S_{n-2}$ diagram appears twice via the Littlewood-Richardson multiplication rules (see footnote \ref{foot}).}\label{f:decompose-2}
 \end{center}
\end{figure}

\begin{figure}
  \begin{center}
 \epsfysize=1.2in
   \mbox{\epsfbox{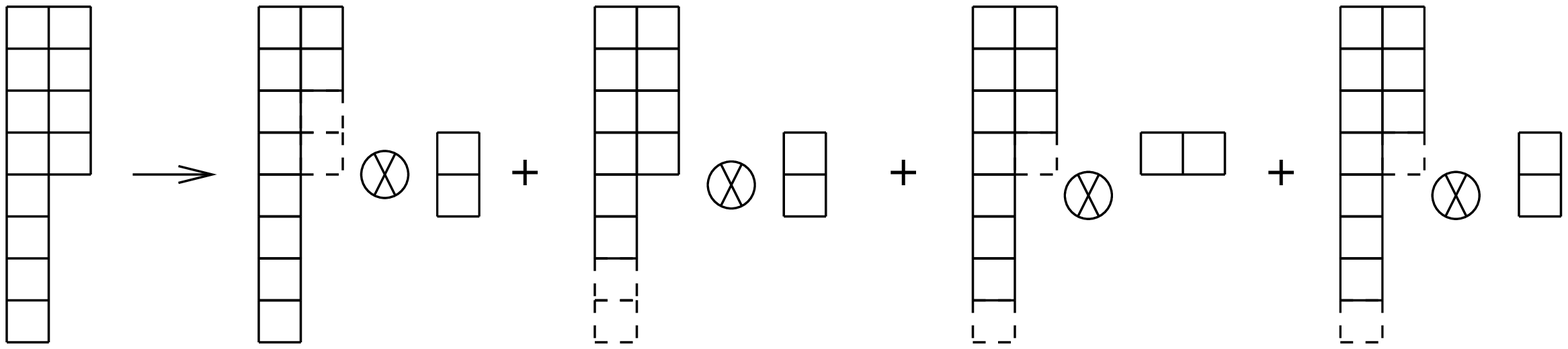}}
    \caption{The decomposition  of $S_n\to S_{n-2}\times S_2$.  The dashed boxes indicate where the boxes of the $S_2$ diagram can attach (see footnote \ref{foot}).}\label{f:decompose-1}
 \end{center}
\end{figure}

With these preparations we can calculate the number of independent operators.  First, note 
that (\ref{action-1}) simply expresses the fact that $A^R_{ij}$ transforms as a tensor product of $D^R$ and $\O {D^R}$. To see this,
assume vector space $V$ with basis $e_1,..., e_l$
carries the representation $R$ and  that a vector space $W$ with basis 
$\eta_1,..., \eta_k$ carries the representation $S$.  The
tensor space $V\otimes W$ with basis $e_i\otimes \eta_{\a}$ satisfies
\bea
g( e_i\otimes \eta_{\a}) & = & (g e_i)\otimes (g \eta_{\a})
 =   \left(e_j D^{R}(g)_{ji}\right) \otimes \left(
\eta_{\b} D^{S}(g)_{\b\a}\right)\nonumber \\
& = & (e_j \otimes \eta_{\b})D^{R}(g)_{ji}D^{S}(g)_{\b\a} 
\Label{action-2}
\eea
Comparing (\ref{action-1}) and (\ref{action-2}) we reach our
conclusion.

This reduces the problem of finding $G\subset S_n$
invariant operators  to {\sl counting the   trivial representations of $G$ inside the
tensor product of $R$ and $\O R$.} This is easy to do.
The group character $\chi$ has a decomposition
\bean
\chi^{R}|_G=\sum_{k} p_k \chi^{k}
\eean
where $R$ is an irreducible representation of $S_n$, 
$k$  is an irreducible representation of $G$ decomposed from 
$R$ and $p_k$  is the number of times the representation $k$ appears in the decomposition. 
Then
\bean
\chi^{\O R}=\sum_{k} p_k\chi^{k*}
\eean
and finally
\bea
\dim & = & \braket{ 1, \chi(R) \chi(\O R)} \nonumber 
 = \sum_{i,j} p_i p_j\braket{ 1, \chi^{i} \chi^{j*}}  \nonumber\\
& = & \sum_{i,j} p_i p_j\braket{ \chi^{j},\chi^{i}} 
 =  \sum_{k} p_k^2 \Label{basis}
\eea

\subsubsection{Examples: Gauss' Law recovered}

We will demonstrate how our general result (\ref{basis}) manifests the constraints of Gauss' Law by considering several examples.

\paragraph{1. No open string excitations: }  If no strings are attached to a 1/2-BPS state of D-branes, we find the operator by  replacing $1,2,...,n\to X$ in the $n$ boxes of a tableau.   The number of independent states thus equal the number of $S_n$ invariant Young elements.  Since each Young diagram corresponds to a single irreducible representation,  there is unique $S_n$ singlet Young element, and corresponding D-brane state.  This is the result of \cite{CJR}.

\paragraph{2. One string attached to two different D-branes: }  This operator is computed by replacing
 $1,2,...,(n-1)\to X$ and $n\to Y$ in a Young diagram with two columns of different lengths.  Fig.~4 showed that the corresponding $S_n$ representation decomposes into a sum of two irreducible $S_{n-1} \times S_1$ representations each with multiplicity 1.  Thus (\ref{basis}) shows that there are two   independent operators.  This matches Gauss' law: a single string starting on a spherical brane must also end on it, and therefore the two branes should give rise to two independent states.  In particular, Gauss' law prohibits  an open string connecting the two branes.

\paragraph{3. One string attached to two identical D-branes: }  In this case, Gauss' law and the equivalence of the two D-branes together imply that there is only one gauge invariant open string state.
Indeed, for this special case the $S_n$ representation with two equal columns decomposes into a single $S_{n-1}$ irreducible representation in which the second column has one less box.    Thus there is only one independent state, as expected.

\paragraph{4. Two distinct open strings attached to two different D-branes: }  We construct the operator by replacing $1,2,...,(n-2)\to X$, $(n-1)\to Y_1$ and $n\to Y_2$ in Young diagram of $S_n$ with two unequal columns.   Fig.~5 shows that the $S_n$ diagram decomposes into a sum of three irreducible $S_{n-2} \times S_1 \times S_1$ representations.  According to (\ref{basis}) this leads to $1^2 + 1^2 + 2^2 = 6$ independent states.   This matches the Gauss law constrained number of states of two distinguishable strings on two distinguishable branes A and B:  both strings on brane A or B (2 states), one string on A and the other on B (2 states), one oriented string stretched from A to B and vice versa (2 states).

\paragraph{5. Two distinct open strings attached to two identical D-branes: }  We construct the operator by replacing $1,2,...,(n-2)\to X$, $(n-1) \to Y_1$ and $n\to Y_2$ in a Young diagram of $S_n$ with two equal columns.  It  is easily seen that in this case the $S_n$ diagram decomposes into a sum of two irreducibles of $S_{n-2} \times S_1 \times S_1$, each with multiplicity 1.   This implies two independent states.   Naively one might have expected three independent states: (i) both strings on a single brane, (ii) one string on each brane, and (iii) both strings stretched between the branes.   However, since the the branes are identical the third state is gauge equivalent to the second so there should be only two independent states.

\paragraph{6. Two identical open strings attached to two different D-branes: }  We construct the operator by replacing $1,2,...,(n-2)\to X$, $(n-1) \to Y_1$ and $n\to Y_1$ in a Young diagram of $S_n$ with two unequal  columns.   As seen in Fig.~6, the $S_n$ diagram decomposes into a sum of four irreducibles of $S_{n-2} \times S_2$, each with multiplicity 1.  This leads by (\ref{basis}) to four independent states.  The expected counting agrees: (a) both strings on a single brane (2 states), (b) one string on each brane (1 state), and (c) both strings stretched between the branes (1 state).

\paragraph{7. Two identical open strings attached to two identical D-branes: }    We construct the operator by replacing $1,2,...,(n-2)\to X$, $(n-1) \to Y_1$ and $n\to Y_1$ in a Young diagram of $S_n$ with two equal  columns.   Unlike Fig.~6, since the two columns of the $S_n$ diagram are equal length, it can be shown that the decomposition into $S_{n-2} \times S_2$ only leads to two irreducible diagrams.  (In effect only the first and third products of Fig.~6 survive.)   Thus there are two independent states.   This is as expected from the discussion in example 5:  all states are gauge equivalent to either two strings on the same brane, or on different branes.

Our proposal for constructing operators describing open string states on D-branes exactly reproduces the expected constraints imposed by Gauss' law.     Remarkably, the combinatorics of Young diagrams even ``knows'' about the distinctions between identical and distinguishable brane and string states.  This is strong evidence  in favor of our  identification between
Young diagram operators and open string states on a system of D-branes.  Although all the examples we have given involve two column Young diagrams, our proposal can be applied in general, for example, to n-column or n-row diagrams.  In other words, our proposal describes the excited states of a general system of D-branes some of which might be extended into $\ads{5}$ or into $S^5$ in the string theory description.

\subsection{Constructing Young operators}

In practical calculations one needs a convenient way of writing down the operators described in the previous section.  Of course, we could always return to the definition (\ref{Young-op}) but this expression is unwieldy to compute.   Below we develop two other techniques for constructing the operator

\paragraph{Projection operator method: }   Recall again that the Young diagram associated to a system of D-branes is also related to an irreducible representation $R$ of $S_n$ where $n$ is the number of boxes in the diagram.  As we have seen, identifying the open string operators on the D-branes requires extracting the trivial representations of some $G \in S_n$ within the tensor product $R \otimes \O R$.  A standard way of doing this is by constructing the appropriate projection operator.   For  the trivial representation the projection operator is 
\bea 
P_{I}={1\over n_G} \sum_{g\in G} T(g)\Label{P-our-1}
\eea
where $T(g)$ is the action of $g$ in the represention $R \otimes \O R$ and $n_G$ is the number of elements in $G$.  Using this, the action of the projection operator on the Young basis elements (\ref{A-def}) is 
\bean
P_I \cdot A_{ij} & = & {1\over n_G} \sum_{g\in G} T(g)\cdot  A_{ij} \\
& = & {1\over n_G} \sum_{g\in G}
\sum_{k,l} A_{kl} D(g)_{ki} \O {D(g)_{lj}} \\
& = & {1\over n_G} \sum_{g\in G}\sum_{k,l} \sum_{s\in S_n}
\O {D_{kl}(s)}  D(g)_{ki} \O {D(g)_{lj}} \cdot s
\eean
where we used (\ref{action-1}) and then (\ref{A-def}). To go further and simplify the
calculation, we can choose a basis so that  the representation $R$ of $S_n$ has block diagonal
form when it is restricted to $G\subset S_n$.  It is convenient to split the index $k$ of the $S_n$ representation $R$ to reflect the decomposition into representations of $G$.   Each such representation of G, which we label $\mu$, could appear several times in $R$, so we will introduce $\alpha_\mu$ to distinguish the multiple appearances.  Then we write   $k=[\mu,\a_\mu,t]$  where in addition to  $\mu$ and  $\a_\mu$ we have included $t = 1 \cdots \dim(\mu)$.   Likewise we split
$i = [\tilde\mu,\tilde\alpha_\mu,\tilde{t}]$ and $i = [\tilde\nu,\tilde\alpha_\nu,\tilde{d}]$.   With this notation,
\bean
& & \sum_{g\in G} \sum_{k,l}D(g)_{ki} \O {D(g)_{lj}}\O {D_{kl}(s)}\\
 & = &
\sum_{g\in G}\sum_{\mu,\nu}\sum_{\a_\mu,a_\nu}\sum_{t,d}
D^{\mu,a_\mu; \W \mu, \W a_\mu}(g)_{t\W t} 
\O {D^{\nu,a_\nu; \W \nu,\W \a_\nu}(g)_{d\W d}}
~~\O {D(s)_{[\mu,\a_\mu,t][\nu,\a_\nu,d]}}\\
& = & \sum_{\mu,\nu}\sum_{\a_\mu,a_\nu}\sum_{t,d} 
{n_G\over n_\mu}
\delta_{\mu\W \mu} \delta_{\nu\W \nu}\delta_{\a_\mu \W \a_\mu}
\delta_{\a_\nu \W\a_\nu}
\delta_{\mu\nu}\delta_{td}\delta_{\W t \W d}
~~\O {D(s)_{[\mu,\a_\mu,t][\nu,\a_\nu,d]}}\\
& = &  {n_G\over n_{\W\mu}}\delta_{\W\mu \W\nu}\delta_{\W t \W d}
\sum_{t}\O {D(s)_{[\W\mu,\W \a_\mu,t][\W\nu,\W \a_\nu,t]}}
\eean
where in the third line we have used the block diagonal form of $D(g)$ and
the corresponding orthogonality relationship (\ref{orth-D}) for $G$.
Assembling everything we have
\bea
P_I \cdot A_{ij} 
& = & \sum_{s\in S_n} {1\over  n_{\W\mu}} 
\delta_{\W\mu \W\nu}\delta_{\W t \W d}
\sum_{t}\O {D(s)_{[\W\mu,\W \a_\mu,t][\W\nu,\W \a_\nu,t]}}\cdot s
\Label{PI-action}
\eea
Nonzero contributions to this projection require
$\W \mu=\W\nu$ and $\W t=\W d$, but $\W \a_\mu,\W \a_\nu$ 
can be arbitrary.  In other words, the identity representation is extracted from the tensor products of any of the multiple occurrences of each representation $\mu$ of $G$ in the decomposition of $R$ with their conjugates.

The expression (\ref{PI-action}) is a weighted sum of permutations.  Applying it to a sequence $\Phi_1 \cdots \Phi_n$ of unitary matrices as in (\ref{Young-op}) we get the desired operator.\footnote{By this we mean that the weighted sum of permutations of lower indices in (\ref{Young-op}) should be replaced by the weighted sum of permutations in (\ref{PI-action}). \label{foot2}}  We use this expression in Sec.~5.4 and the appendices to show orthogonality for 1-string states on arbitrary D-brane states.

\paragraph{Matrix method: }
In fact,  (\ref{PI-action}) can be understood directly from
matrix algebra. For example, suppose $R\to R_1+2 R_2$ under the decomposition
$S_n\to G$.   We can choose a basis for $R$ so that
\bean
D^{R}(g)=\left[ \begin{array}{ccc} D^{R_1} & 0 & 0 \\ 0 & D^{R_2} & 0\\
0 & 0 &  D^{R_2} \\ \end{array} \right],~~~~~\forall g\in G
\eean
Then using the middle form of (\ref{action-1}) we have
\bean
A_{ij} & \to & \left[ \begin{array}{ccc} \O {D^{R_1}}^{-1} & 0 & 0 \\ 
0 & \O {D^{R_2}}^{-1} & 0\\
0 & 0 &  \O {D^{R_2}}^{-1} \\ \end{array} \right] \cdot A_{ij}
\cdot \left[ \begin{array}{ccc} \O {D^{R_1}} & 0 & 0 \\ 0 & 
\O {D^{R_2}} & 0\\
0 & 0 & \O  {D^{R_2}} \\ \end{array} \right] \\
& = & \left[  \begin{array}{ccc} \O {D^{R_1}}^{-1} A_{11}
 \O {D^{R_1}}~~~ & \O {D^{R_1}}^{-1} A_{12}\O {D^{R_2}}~~~ &
\O {D^{R_1}}^{-1} A_{13}\O {D^{R_2}}~~~ \\
\O {D^{R_2}}^{-1} A_{21}  \O {D^{R_1}}~~~ & 
\O {D^{R_2}}^{-1} A_{22}  \O {D^{R_2}}~~~ &
\O {D^{R_2}}^{-1} A_{23}  \O {D^{R_2}}~~~ \\
\O {D^{R_2}}^{-1} A_{31}  \O {D^{R_1}}~~~ & 
\O {D^{R_2}}^{-1} A_{32}  \O {D^{R_2}}~~~ &
\O {D^{R_2}}^{-1} A_{33}  \O {D^{R_2}}~~~ \\
\end{array} \right]
\eean
where we introduced block indices $1,2,3$.  Then it is evident that traces of some of the blocks, namely
 $\Tr A_{11}$, $\Tr A_{22}$, $\Tr A_{23}$, 
$\Tr A_{32}$ and  $\Tr A_{33}$, provide a basis of $G$-invariant group algebra elements.   This is simply an explicit realization of (\ref{PI-action}) that can be used to construct gauge-invariant Yang-Mills operators as described in footnote (\ref{foot2}).

It is useful to illustrate these methods with a couple of examples.  First consider a general 1/2-BPS state.   Then using  (\ref{A-def}), the $S_n$ invariant element of the group algebra is  ${\cal O}=\sum_{i=1}^d A_{ii}=\sum_{s\in S_n} \chi^R(s) s$.  (Here we have used the property $\chi(s)=\chi^*(s)$ 
because all irreducible representations of $S_n$ have real
characters.)  This reproduces the Schur polynomical form given in \cite{CJR}, and the corresponding gauge invariant operator can be constructed as described above.  Second, consider a system of two distinct  D-branes and an open string.   Then the representation $R$ of $S_n$ decomposes to a sum  $R_1+R_2$ of $S_{n-1}$ as described in the examples in the previous section.  The two associated $S_{n-1}$ invariant group algebra elements  are
\bea
{\cal O}_{R,R_1} & = & \sum_{s\in S_n} \sum_{i=1}^{d_{R_1}} 
\O D^R_{ii}(s) s \Label{One-open-1}\\
{\cal O}_{R,R_2} & = & \sum_{s\in S_n} \sum_{i=d_{R_1}+1}^{d_{R}} 
\O D^R_{ii}(s) s \Label{One-open-2}
\eea
where we have chosen $D^R(g\in S_{n-1})$ to be block diagonal.  It is not clear
how to simplify (\ref{One-open-1}) and  (\ref{One-open-2}) further.    While for 
$g\in G$, the block trace in these operators is simply the charcater of $R_1$ and $R_2$, for
 $g\in S_n, g\not\in G$ there is no obvious clean interpretation.
It would be nice to have a more powerful technique for evaluating these traces.

\subsection{Tree level orthogonality of Young operators}

The combinatorial emergence of Gauss's law from our Young diagrams is strong evidence that we have correctly identified the operators describing strings on D-branes from the Yang-Mills perspective.  However, it would also be nice to prove that we have a good orthogonal basis, at least at tree level and in the large $N$ limit.   The challenge here is that we do not have a general expression for the block traces the previous section, and a case by case enumeration of the multi-string operators is not tractable.  However, we have been able to show that the various operators describing single strings attached to a multi-brane system are indeed orthogonal.  This amounts to showing that the matrix of 2-point functions of these operators is diagonal.   The general calculation is lengthy and is presented in Appendix C.
The result is
\bea
\braket{{\cal O}_{R,R_1}^\dagger {\cal O}_{S,S_1}}\sim 0,~~~R\neq S~~or~~R_1\neq 
S_1 \Label{one-bases}
\eea
in large $N$ limit.  Here $R,S$ are $S_n$ representations that determine the system of branes and $R_1,S_1$ are $S_{n-1}$ representations that determine the open string state.  This shows that at least arbitrary combinations of 1/2-BPS D-branes with one string on them, our proposal provides a good orthogonal basis of operators.

\section{Dynamical emergence of non-Abelian structure}

In the previous section we took the first step towards demonstrating the appearance of a new gauge symmetry by showing how Gauss' law emerges from the combinatorics of Young diagrams.   In this section we will provide evidence that the emergent symmetry will be non-Abelian in the presence of multiple branes by showing how perturbative loops sense the rank of the group.  From the perspective of string theory we will show that the Chan-Paton factors associated with strings in a multi-brane system dynamically emerge from Yang-Mills theory.  Our basic tactic is to compute the one-loop anomalous dimension of operators representing a string attached to two branes.   When the branes are identical we should expect a result that is twice the answer for a single brane because of the trace over the indices of  the enhanced gauge group associated to coincident branes, or, equivalently,  because open strings can end on multiple branes (see Fig.~8).

The operator expressions (\ref{PI-action}) are still too complicated for a general calculation.  Hence we will consider a special case: an open string attached to a two D-brane system in which one brane is of maximal size.    Gauss' law will forbid a string stretched between two such branes, so there should two states: a string attached to the small brane and a string attached to the big  brane.    The obvious guess for the corresponding operators is
\bea
P_1 &\equiv &  \epsilon^{i_1 i_2 ... i_N }_{j_1 j_2 ... j_N }
\Phi^{j_1}_{i_1}  \Phi^{j_2}_{i_2}... \Phi^{j_N}_{i_N}
 \epsilon^{\W i_1 \W i_2 ...\W i_M \W k_1 ...\W k_{N-M}}_
{\W j_1 \W j_2... \W j_M \W k_1 ... \W k_{N-M}}
\Phi^{\W j_1}_{\W i_1} \Phi^{\W j_2}_{\W i_2}... 
\Phi^{\W j_{M-1}}_{\W i_{M-1}} (Y^{J})^{\W j_M}_{\W i_M} \Label{P11}
\\
P_2 &\equiv & \epsilon^{i_1 i_2 ... i_N }_{j_1 j_2 ... j_N }
\Phi^{j_1}_{i_1}  \Phi^{j_2}_{i_2}... \Phi^{j_{N-1}}_{i_{N-1}}
(Y^{J})^{ j_N}_{ i_N}
 \epsilon^{\W i_1 \W i_2 ...\W i_M \W k_1 ...\W k_{N-M}}_
{\W j_1 \W j_2... \W j_M \W k_1 ... \W k_{N-M}}
\Phi^{\W j_1}_{\W i_1} \Phi^{\W j_2}_{\W i_2}... 
\Phi^{\W j_{M}}_{\W i_{M}}\Label{P22}
\eea
We could naively interpret (\ref{P11}) as the product of an operator describing a maximal size D-brane ($\det\Phi$) and an operator describing a smaller D-brane (${\rm subdet}\Phi$) with a string ($Y^J$) attached to the latter.  Similarly (\ref{P22}) is naively a product of operators describing a maximal brane with a string on it, and a smaller brane.  However, as shown in Appendix~B this naive guess is not quite correct -- the two operators make states that are not orthogonal to each other.   Rather, following the  proposal in the previous section the two orthogonal one string states should be described by the Young diagrams in Fig.~7.     The corresponding operators are:
\bea 
{\cal O}_{1}& =& { (N+1)\over (N-M+1)} P_1\Label{Young-11}\\
{\cal O}_{2} & = & {N\over (N-M)} P_2 -{ M\over (N-M)(N-M+1)}P_1
 \Label{Young-22}
\eea
Following the calculations in Appendix~B, $\CO_1$ and $\CO_2$ make orthogonal states at tree level.   It is pleasant to note that when $M \ll N$ so that the two branes are very different in size, the naive guess is correct, i.e., $\CO_i \approx P_i$.

\begin{figure}
\Label{stringontwobranes}
  \begin{center}
 \epsfysize=1.2in
   \mbox{\epsfbox{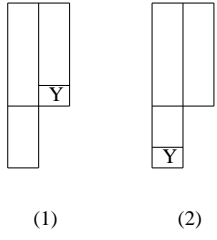}}
    \caption{The two orthogonal Young diagrams representing a string ($Y$) attached to a pair of D-branes of different sizes}
 \end{center}
\end{figure}

How small should $M$ be so that the branes are well separated?   Recall on $\ads{5} \times S^5$ a single brane with R-charge $M$ moves on a circle of radius
\begin{equation}
r = R \left( 1 - {M \over N} \right)^{1/2} \, ,
\end{equation}
where $R$ is the AdS scale.  The large brane (with $M = N$) is thus at the origin $r = 0$.  In order for the branes to be separated by more than a string length we require
\begin{equation}
r > l_s \, .
\end{equation}
Recalling that $R \sim l_s \lambda^{1/4}$ where $\lambda$ is the 't Hooft coupling, and we are dropping factors $2\pi$, this condition translates into
\begin{equation}
M < N \left( 1 - {1 \over \sqrt{\lambda} }\right)
\Label{specbound}
\end{equation}
So we if we take $M = \alpha N$ for any $\alpha < 1 - 1/\sqrt{\lambda}$ the branes are separated by more than a string length at large 't Hooft coupling and the spectator brane should be largely irrelevant to the dynamics of a string on one of the branes.    However, in the perturbative calculations below we work at small 't Hooft coupling for which this bound is very restrictive.   When $\lambda$ is very small, the scale of both $\ads{5}$ and $S^5$ is the string length, so the branes are never much more than a string length apart. 

Weak coupling results for BPS observables can be extrapolated to strong coupling where the $S^5$ is large and the supergravity analysis is valid, but the operators (\ref{Young-11},\ref{Young-22}) are only near-BPS.    Indeed, we will see leading order agreement with the supergravity expectations, but there will be small $O(1)$ differences.  Indeed the fact that (\ref{specbound}) contains a $\sqrt{\lambda}$ immediately tells us that a naive perturbative large N analysis in the Yang-Mills theory will not directly reproduce this bound.

To arrive at our results below we will be computing correlation functions of operators like $P_1$ and $P_2$.  In these calculations, while the contractions between the $\Phi$s will be carried out exactly, those between the $Y$s are done at the planar level.  In this sense our computations are at the leading order in large $N$.   One loop corrections from interactions are also evaluated in the planar limit for contractions between $Y$s.  The latter are suppressed by powers of the coupling, but not by powers of $N$.

\subsection{Rank of the emergent gauge group}

We will now show that as the two branes (\ref{Young-11}) and (\ref{Young-22}) coincide, the one-loop anomalous dimension of an open string becomes twice the answer for a single brane.  This demonstrates the dynamical emergence of the Chan-Paton factors for open strings on coincident branes, or, equivalently, the enhancement of the low energy gauge theory to a non-Abelian group.   The calculations are straightforward but lengthy.  Hence we will not present the detailed steps,\footnote{Appendix~D gives a flavor of the computations involved.} but present the results directly.  We will always assume that $M$ is $O(N)$, so that the smaller object can be interpreted as a D-brane rather than as a closed string ($M \sim O(\sqrt{N})$) or as a supergravity mode ($M \sim O(1)$).

\paragraph{String on the large brane: }  The general expressions simplify when $(M - N)$ is $O(N)$ also.   This condition is equivalent to setting 
\begin{equation}
M = \alpha N   ~~~~;~~~~ 0 < \alpha < 1  
\end{equation}
so that 
\begin{equation}
\CO_2 \propto P_2
\end{equation}
to leading order in $N$.  The tree level and 1-loop two-point functions of $P_2$ turn out to be
\begin{eqnarray}
\langle P_2(x) \bar{P}_2(0) \rangle_{{\rm tree}} &=&
{N^{J-1} (N-M)!N!^4 M!^2 g_s^{N+M+J-1} \over (N-M + 1) (2\pi |x|^2)^{N+M+J-1}} \equiv 
{C_2 \over |x|^{2\Delta}} \Label{P2-tree}\\
\langle P_2(x) \bar{P}_2(0) \rangle_{{\rm 1-loop}} &=&
- {2 g_s \over \pi} (J - 1) (1 + {M\over N})  {C_2 \over x^{2\Delta}}   \log(|x| \Lambda)
 \Label{P2-one-loop}
\end{eqnarray}
where $\Delta = N+M+J -1$ is the free field  dimension of the operator. From this we read off the anomalous dimension 
\begin{equation}
\delta\Delta_{{\rm 1-loop},2} = {g_s (J-1) \over \pi} (1 + {M\over N})  
{g_s (J-1) \over \pi} (1 + \alpha)
\, .
\Label{largeoneloopanom}
\end{equation}
When $\alpha \ll 1$, so that the spectator brane is far from the larger one, this approaches the the result for a string on a single maximal sized D-brane ($N=M$ in (\ref{sub})).    By contrast, when $\alpha \to 1$, so that the spectator brane approaches the larger one, we get twice the single brane answer.     This exactly reproduces the expected effect of having multiple coincident branes in open string loop diagrams.   One loop and two loop diagrams for open string propagation are depicted in Fig.~8. 
In the large $N$ limit, such open string loops are identified with loop amplitudes in the gauge theory.    
At  $l$ loops there are $l$ inner circles in Fig.~8, each of which can end on any of the branes present, enhancing the amplitude by $G^l$. Hence the factor of $2$ in the one-loop contribution (\ref{largeoneloopanom}).  At low energies, the dynamics of the string should then give rise to a non-Abelian gauge theory as $\alpha \to 1$.

\begin{figure}
  \begin{center}
 \epsfysize=2.0in
   \mbox{\epsfbox{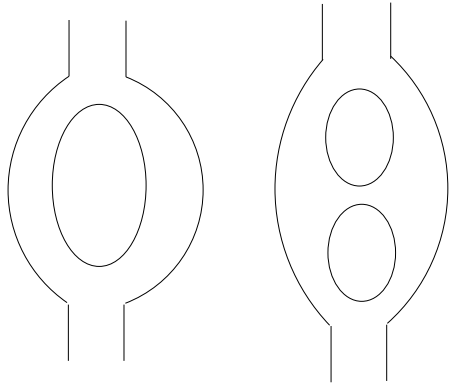}}
\end{center}
\caption{An one loop open string diagram and a two loop open
string diagram. The inner lines (circles) represents the branes
that open string can end on. } \label{figure1}
\end{figure}

\paragraph{String on the small brane: }  In this case we assume $M = O(N)$ and keep terms at leading order in $M$ and $N$.  The tree level and one loop two-point functions are:
\begin{eqnarray}
\langle P_1(x) \bar{P}_1(0) \rangle_{{\rm tree}} &=&
{N^{J} \,(N-M)!\, N!^4 \, M! \, (M-1)! \, g_s^{N+M+J-1} \over  (N-M+2)(2\pi |x|^2)^{N+M+J-1}} 
\equiv 
{C_1 \over |x|^{2\Delta}} \Label{P1-tree}\\
\langle P_1(x) \bar{P}_1(0) \rangle_{{\rm 1-loop}} &=&
- {2 g_s \over \pi} ((N-M) + 2J -1 - {1 \over N - M +1})  {C_1 \over x^{2\Delta}}   \log(|x| \Lambda)
 \Label{P1-one-loop}
\end{eqnarray}
where $\Delta$ is the free field dimension of the operator.  The anomalous dimension is therefore
\begin{equation}
\delta\Delta_{{\rm 1-loop},1} = {g_s \over \pi} ((N - M) + 2J - 1 - {1 \over N - M + 1})
\Label{smalloneloopanom}
\end{equation}
As $M \to N$, like (\ref{largeoneloopanom}), this approaches twice the answer for a single maximal size brane, reproducing the expected enhancement due the presence of multiple branes.   When $M \ll N$ the spectator maximal size brane is far from the brane and (\ref{smalloneloopanom}) agrees at leading order with the result for a string on a single brane ($M < N$ in (\ref{sub})).   At the next to leading order, there is a disagreement by a factor of 2 in front of $J$.   It is expected that there will be some disagreement.  We are extrapolating from very weak 't Hooft coupling where the separation is never more than a string length to strong coupling where the branes are well separated.  Since we are not discussing BPS quantities, it is surprising that the agreement is so good at the subleading level.  The near-BPS character of the operators is conferring some residual protection to the anomalous dimension.

\paragraph{Discussion: }  As $M \to N$, we should strictly speaking account for the the non-vanishing overlap between $P_2$ and $P_1$ by taking the orthogonal combinations (\ref{Young-11},\ref{Young-22}).    We have not done so because when $M = N$, in fact $P_2 = P_1$ and all overlaps give the same answer which will be twice the answer for a single brane as shown above.   When $M \ll N$ we can likewise neglect the overlaps of $P_2$ and $P_1$ at the present order of analysis, but for intermediate values of $M$ it is important to correctly account for the tree-level diagonalization of operators.   In any case, our results clearly show the emergence of the Chan-Paton factors associated to multiple coincident branes.    At low energy this should lead to a new non-Abelian gauge symmetry arising from the brane dynamics. As the branes separate (i.e., as $M$ is changed), the Chan-Paton factors discretely interpolate between integers.

\subsection{Emergent string interactions}

In previous sections we established that: (1) open strings emerge from the dynamics of D-brane operators in Yang-Mills theory,  (2) Gauss' law is realized on multi-brane systems, (3) the Chan-Paton factors of multiple branes appear dynamically, plausibly allowing the appearance of non-Abelian gauge theory on multiple coincident branes.  This is enhanced by the fact that we seem to get the correct counting of single string and two string states for the non-abelian theory.

The challenge now is to establish that the excitations of the D-branes  have the detailed interactions of  a (emergent) gauge theory.    Here we take the first step by simply displaying the kinds of CFT calculations that should be related to the splitting and joining of open strings on D-branes.

\subsubsection{Interactions of strings in their ground state}
Let us study the basic interaction of string theory -- the splitting of one string into two.    The simplest situation to examine is one in which the open strings are in their ground states.  Operators corresponding to D-branes with one and two string excitations are:
\begin{eqnarray}
\CO_1 &=& \epsilon_{i_1\cdots i_M}^{j_1\cdots j_M}Z^{i_1}_{j_1}\cdots
Z^{i_{M-1}}_{j_{M-1}}(Y^J)^{i_M}_{j_M} \\
\CO_2&=&\epsilon_{i_1\cdots i_{M+1}}^{j_1\cdots
j_Mj_{M+1}}Z^{i_1}_{j_1}\cdots
Z^{i_{M-1}}_{j_{M-1}}(Y^{J_1})^{i_M}_{j_M}(Y^{J_2})^{i_{M+1}}_{j_{M+1}}
\end{eqnarray}
Here $Y=\frac{1}{\sqrt{2}}(\phi^3+i\phi^4)$ and the R-charge
$J=J_1+J_2$ is large, but $1 \ll J \ll M,N$.  (We will typically take $J\sim \sqrt{N}$ leading to a string with well-controlled interactions of the kind studied in \cite{BMN,BHLN}.)\footnote{Recall that a string with spacetime mass $E \sim 1/l_s$ has a CFT dimension $\Delta \sim R/l_s \sim N^{1/4}$ where $R$ is the AdS scale, and a near BPS state should have R-charge $J \sim \Delta$.}  The two point
function $\langle \CO_1 \overline{\CO}_2 \rangle$ will be related to the 
open string field theory vertex describing an open string on a giant
graviton splitting into two open strings, as in the case
of BMN operators in \cite{Huang, SV}.   Open string field
theory in a context related to the present one was studied in \cite{Open}. 

For simplicity we only
consider the free field correlator $\langle \CO_1 \overline{\CO}_2
\rangle$. We also omit the spacetime dependent factor
$\frac{g_s}{2\pi|x|^2}$ to simplify notation.  To leading order in $N$, the normalizations and
interactions of our operators are
\begin{eqnarray}
\langle \CO_1\overline{\CO}_1 \rangle &=& 
\frac{(M-1)!^2(M-1)(N-2)!N^{J+1}}{(N-M)!} \, ,\\
\langle \CO_2\overline{\CO}_2 \rangle &=&
\frac{(M-1)!^2(M-1)(M-2)(N-2)!N^{J}}{(N-M-1)!} \, ,\\
\langle \CO_1\overline{\CO}_2 \rangle &=& -2
\frac{(M-1)!^2(M-1)(N-2)!N^{J}}{(N-M-1)!} \, ,
\end{eqnarray}
leading to the normalized interaction amplitude
\footnote{In computing $\langle
\CO_2\overline{\CO}_2 \rangle$ we have assumed $J_1\neq J_2$.
$\langle\CO_2\overline{\CO}_2 \rangle$ will have an extra factor of
$2$, if $J_1=J_2$ .}
\begin{equation} 
\frac{\langle \CO_1\overline{\CO}_2 \rangle}{\sqrt{\langle
\CO_1\overline{\CO}_1 \rangle \langle \CO_2\overline{\CO}_2
\rangle}}=-2\sqrt{\frac{N-M}{(M-2)N}} \, .
\Label{77}
\end{equation}
Interestingly, the amplitude is independent of the R-charges.     When both $M$ and $N$ are of $O(N)$, the amplitude is $O(1/\sqrt{N})$.     This is suppressed at large $N$.   This may seem surprising given that at large $J$ we have a very heavy string state.  In fact, such suppression of the decay amplitude of heavy rotating rotating open strings occurs even  for D-branes in flat space \cite{vijayigor}.  (Similar results for closed strings are obtained in \cite{iengorusso}.)   Nevertheless,  if the R-charge $J$ is  large (say $O(\sqrt{N})$), the sum over partitions of $J$ into $J_1$ and $J_2$ will lead to an unsupressed inclusive splitting amplitude.  This is reminiscent of the closed string splitting amplitude  described by BMN operators \cite{Huang}.

\paragraph{Closed string emission: } We can also study amplitudes for closed string emission from an excited D-brane.  The operator corresponding to a D-brane along with an emitted closed string is
\begin{equation}
\CO_3=\epsilon_{i_1\cdots i_{M-1}}^{j_1\cdots
j_{M-1}}Z^{i_1}_{j_1}\cdots Z^{i_{M-1}}_{j_{M-1}}\tr(Y^J)
\end{equation}
At leading order in $N$ the two point functions are
\begin{equation}
\langle \CO_3\overline{\CO}_3 \rangle=\frac{
J(M-1)!^2N!N^{J}}{(N-M+1)!} ~~~~;~~~~
\langle \CO_1\overline{\CO}_3 \rangle= \frac{J
(M-1)!^2(N-1)!N^{J}}{(N-M)!}
\end{equation}
leading to the normalized interaction amplitude
\begin{equation} 
\frac{\langle \CO_1\overline{\CO}_3 \rangle}{\sqrt{\langle
\CO_1\overline{\CO}_1 \rangle \langle \CO_3\overline{\CO}_3
\rangle}}= \sqrt{\frac{(N-M+1)J}{(M-1)N}}
\Label{open-close}
\end{equation}
The amplitude scales as $\sqrt{\frac{J}{N}}$, if $M$ and $N$ are of 
the same order.

\subsubsection{Splitting and joining of excited string states}
In Sec.~4 we showed that (\ref{opNeumann}) and (\ref{opDirichlet}) are open string fluctuations that realize Neumann and Dirichlet boundary conditions respectively.  We can consider the interactions of such excited string states.   Following \cite{BMN, BHLN}, it is convenient to ``Fourier transform'' the operators (\ref{opNeumann},\ref{opDirichlet}) representing single oscillator excitations with Dirichlet or Neumann boundary conditions to get:
\begin{eqnarray}
\CO_1^{D,m}&=& \epsilon_{i_1\cdots i_M}^{j_1\cdots
j_M}Z^{i_1}_{j_1}\cdots Z^{i_{M-1}}_{j_{M-1}}(\sum_{l=0}^{J} \sin
(\frac{\pi ml}{J})  Y^l\phi^I Y^{J-l})^{i_M}_{j_M} \\
\CO_1^{N,m}&=&\epsilon_{i_1\cdots i_M}^{j_1\cdots
j_M}Z^{i_1}_{j_1}\cdots Z^{i_{M-1}}_{j_{M-1}}(\sum_{l=0}^{J} \cos
(\frac{\pi ml}{J})  Y^l\phi^K Y^{J-l})^{i_M}_{j_M}
\end{eqnarray}
Here $D$ represents Dirichlet boundary conditions and $\phi^I=\phi^5, \phi^6$, while $N$ represents Neumann boundary conditions with $\phi^K=\phi^1, \phi^2$ \cite{BHLN}.  The worldsheet momentum $m$ satisfies $0\leq m<J$ for Neumann modes and $0<m<J$ for Dirichlet modes.  

Similarly, operators representing a brane with two strings, one of which is excited by an oscillator mode, are given by
\begin{eqnarray}
\CO_2^{D,m} &=& \epsilon_{i_1\cdots i_Mi_{M+1}}^{j_1\cdots
j_Mj_{M+1}}Z^{i_1}_{j_1}\cdots
Z^{i_{M-1}}_{j_{M-1}}(\sum_{l=0}^{J_1} \sin (\frac{\pi ml}{J})
Y^l\phi^I Y^{J_1-l})^{i_M}_{j_M}(Y^{J_2})^{i_{M+1}}_{j_{M+1}} \\
\CO_2^{N,m} &=& \epsilon_{i_1\cdots i_Mi_{M+1}}^{j_1\cdots
j_Mj_{M+1}}Z^{i_1}_{j_1}\cdots
Z^{i_{M-1}}_{j_{M-1}}(\sum_{l=0}^{J_1} \cos (\frac{\pi ml}{J})
Y^l\phi^K Y^{J_1-l})^{i_M}_{j_M}(Y^{J_2})^{i_{M+1}}_{j_{M+1}}
\end{eqnarray}
Finally, the operator describing a brane and a closed string is
\begin{eqnarray}
\CO_3^{0}=\epsilon_{i_1\cdots i_{M-1}}^{j_1\cdots
j_{M-1}}Z^{i_1}_{j_1}\cdots Z^{i_{M-1}}_{j_{M-1}} \tr (\phi^I Y^J) \, .
\Label{braneclosedagain}
\end{eqnarray}
Because of the level matching condition for closed strings, only a zero mode excitation is possible in (\ref{braneclosedagain}) with a single insertion of $\phi$ into the string of $Y$s \cite{BMN}. 

As in the previous subsection, we calculate the two point
functions of these operators. We find that for string splitting
$\langle \CO_1^{D,m_1}\CO_2^{D,m_2} \rangle$ or $\langle
\CO_1^{N,m_1}\CO_2^{N, m_2} \rangle$, the result is ($J=J_1+J_2$)
\begin{eqnarray}
 \frac{\langle \CO_1^{N(D),m_1}\overline{\CO}_2^{N(D),m_2}
\rangle}{\sqrt{\langle \CO_1^{N(D),m_1}\overline{\CO}_1^{N(D),m_1}
\rangle \langle \CO_2^{N(D),m_2}\overline{\CO}_2^{N(D),m_2} \rangle}}  
 -2 \ \sqrt{\frac{N-M}{(M-2)N}} \ \frac{1}{\sqrt{J_1J}} \ \cos (\frac{\pi
m_1J_2}{2J})\nonumber \\ 
 \times \ \ \ 
\Bigl\{ \ \frac{\sin(\frac{(J_1+1)\pi}{2}(\frac{m_1}{J}-\frac{m_2}{J_1}))}
{\sin(\frac{\pi}{2}(\frac{m_1}{J}-\frac{m_2}{J_1}))}  \ 
\cos(\frac{\pi m_1J_2}{2J}+\frac{\pi
J_1}{2}(\frac{m_1}{J}-\frac{m_2}{J_1})) \ \ \ \ \nonumber \\
\pm \ 
\frac{\sin(\frac{(J_1+1)\pi}{2}(\frac{m_1}{J}+\frac{m_2}{J_1}))}
{\sin(\frac{\pi}{2}(\frac{m_1}{J}+\frac{m_2}{J_1}))} \ 
\cos(\frac{\pi m_1J_2}{2J}+\frac{\pi
J_1}{2}(\frac{m_1}{J}+\frac{m_2}{J_1})) \  \Bigr\}
\end{eqnarray}
Here we assume that $J_1$ and $J$ are both large and of the same order.  The amplitude is suppressed relative to the ground state splitting computed in (\ref{77}) unless $\frac{m_1}{J} \pm \frac{m_2}{J_1}\sim \frac{1}{J}$ or $\frac{m_1}{J}-\frac{m_2}{J_1}\sim \frac{1}{J}$ in which case one or both of the terms within the braces will be large enough to compensate for the overall $1/\sqrt{J_1 J}$.  Notes that all these conditions imply $\frac{m_1}{J} - \frac{m_2}{J_1}\sim \frac{1}{J}$ since $ 0\leq \frac{m_1}{J}, \frac{m_2}{J_1} <1$.   The condition $\frac{m_1}{J}=\frac{m_2}{J_1}$ should be understood as light cone momentum conservation, and is analogous to the level
matching conditions in closed string case. For closed string emission, only the zero mode with Neumann
boundary conditions contributes. The result remains the same as (\ref{open-close}):
\begin{eqnarray}
\frac{\langle \CO_1^{N,0}\overline{\CO}_3^{0}
\rangle}{\sqrt{\langle \CO_1^{N,0}\overline{\CO}_1^{N,0} \rangle
\langle \CO_3^{0}\overline{\CO}_3^{0} \rangle}}\sqrt{\frac{(N-M+1)J}{(M-1)N}}
\end{eqnarray}

\section{Conclusion}

In this paper we have accumulated evidence, building on previous works, that certain 1/2-BPS operators of $\CN$ Yang-Mills theory are D-branes: (1) Their string-like excitations have Neumann and Dirichlet boundary conditions in appropriate directions,   (2) Multi-brane states beautifully realize Gauss' Law for strings ending on them via the combinatorics of associated Young diagrams, (3) The Chan-Paton factors associated with multiple coincident branes emerge dynamically.  We have to temper our optimism of 
the prescription we have given, because
there are some details of the description of strings stretching between D-branes that  we
were not able to do systematically. We have given a list of operators which we claim is a complete solution to the problem of stretching strings between giant gravitons, but we were not able to prove in general that the states corresponding to different configurations are orthogonal. Since we have given the list of operators, it is in principle possible to determine if our prescription is complete or not. We are currently looking into this issue.
We also proposed a second basis for which we could do calculations systematically and we gave an implicit argument that they are equivalent. This was  not at a level where we can translate between the two prescriptions by some sort of Fourier transform. We believe this issue should be explored further.

In Sec.~6 we also showed in this second basis how to compute CFT correlators associated with the interactions of strings on D-branes.   It will be interesting to properly identify  the dictionary relating these computations (e.g., (\ref{77}) and (\ref{open-close})) to the string field theory  on D-branes in $\ads{5} \times S^5$.    It is particularly intriguing that  the low-energy dynamics of the branes should be described by an emergent gauge theory which is local, not on the $S^3$ on which the original Yang-Mills is defined, but rather on a new $3+1$ dimensional space which appears from the matrix degrees of freedom in the Yang-Mills theory.   The emergent locality will involve an exchange between the $SO(4)$ rotation group of the original $S^3$ and the $SO(4)$ subgroup of the R-symmetry group that survives the introduction of a 1/2-BPS D-brane operator.  Evidence from black hole dynamics led \cite{BN} to  propose that in the presence of very heavy D-brane states the low-energy dynamics of a Yang-Mills theory should enjoy  a duality with an emergent field theory which is local on a different space and can have different gauge and global symmetries.\footnote{It was also suggested in \cite{BN} that the new gauge group could have a higher rank than the Yang-Mills theory we start with.  At least at the BPS level this looks problematic in view of the new class of supergravity solutions that has appeared in \cite{linluninmalda}.  However, away from the BPS limit it may yet be possible for a lower rank gauge theory to have dynamics that is effectively described by a higher rank gauge theory.  Indeed, the evidence for this from black hole entropy in \cite{BN} was in a non-BPS setting. See also the recent discussion \cite{Gub,nemani} about how to count the entropy of these states}   The present paper establishes basic ingredients needed for showing how such a duality can arise.

It is worth mentioning that there is a considerable literature on how locality in the {\it bulk} of $\ads{5} \times S^5$ (as opposed to on a D-brane)  arises from the CFT point of view (e.g., \cite{BDHM,BKLT,Gidd,HH,Bgiant}). In some cases like  \cite{PS}, knowing were interactions happen helps to resolve issues between 'field theory' behavior and 'string theory'  behavior in scattering amplitudes. Perhaps even more important than understanding how locality appears, is a detailed description of how it breaks down. This is expected to give us some clue as to what happens at the Planck scale in a theory of quantum gravity, and perhaps will lead to a mathematically rigorous  definition of the 'holographic principle', as opposed to the 'holographic behavior' that we can recognize in the AdS/CFT correspondence.

\paragraph{Acknowledgments}
We thank Oleg Lunin for collaboration in the early stages of this work, and Asad Naqvi, Aki Hashimoto for many useful discussions about this material.   Work on this project at Penn (V.B. and M-x.H.) was supported by  the NSF grant PHY-0331728, by the DOE grant DE-FG02-95ER40893 and by an NSF Focused Research Grant DMS0139799 for ``The Geometry of Superstrings". D. B. work supported in part by DOE grant DE-FG02-91ER40618 and NSF under grant No. PHY99-07949.  B.F. was supported by NSF grant PHY-0070928.  M-x.H. thanks the second Simons workshop in mathematics and physics at Stony Brook for hospitality during part of the works.

\appendix

\section{Some combinatorial properties} 
\label{property}
In this appendix we list some properties of the epsilon tensor that are
useful for our calculations. The epsilon tensor is defined as
\begin{eqnarray}
\epsilon^{i_1\cdots i_p}_{j_1\cdots j_p}=\ \left\{
\begin{array}{cl}
1  & \textrm{if $(i_1\cdots i_p)$ is an even permutation of
$(j_1\cdots j_p)$ };
\\[.1in]
-1 & \textrm{if $(i_1\cdots i_p)$ is an odd permutation of
$(j_1\cdots j_p)$ };  \\[.1in] 0 & \textrm{otherwise.}
\end{array}
\right.
\end{eqnarray}
We work with  $SU(N)$ gauge groups, so $p\leq N$ and both 
$i_1,\cdots, i_p$ and $j_1, \cdots, j_p$ are integers from $1$ to
$N$.   As special cases we write:
\begin{eqnarray}
\epsilon^{i_1\cdots i_N} &=& \epsilon^{i_1\cdots i_N}_{1\cdots N}, \\
\epsilon^{i_1\cdots i_N}\epsilon_{j_1\cdots
j_N}&=&\epsilon^{i_1\cdots i_N}_{j_1\cdots j_N}, \\
\delta^{i}_{j}&=&\epsilon^{i}_j
\end{eqnarray}
Some  properties of the epsilon tensor that are useful in computations of correlation functions of determinant and sub-determinant operators are:
\begin{eqnarray}
 \label{formula1}
\epsilon^{i_1\cdots i_p}_{j_1\cdots
j_p}&=&\sum_{x=1}^{p}(-1)^{x+1}\delta^{i_1}_{j_x}\epsilon^{i_2\cdots
i_p}_{j_1\cdots j_{x-1}j_{x+1}\cdots j_p} \\
 \label{formula2}
\epsilon^{i_{p+1}\cdots i_Mk_{M+1}\cdots k_N}_{i_{p+1}\cdots
i_Mj_{M+1}\cdots j_N}&=&\frac{M!}{p!}\epsilon^{k_{M+1}\cdots
k_N}_{j_{M+1}\cdots j_N} \\
 \label{formula3}
\epsilon^{k_{M+1}\cdots k_N}_{k_{M+1}^{\prime}\cdots
k_N^{\prime}}\epsilon^{i_{p+1}\cdots i_M k_{M+1}^{\prime}\cdots
k_N^{\prime}}_{j_{p+1}\cdots j_M l_{M+1}\cdots
l_N}&=&(N-M)!\epsilon^{i_{p+1}\cdots i_M k_{M+1}\cdots
k_N}_{j_{p+1}\cdots j_M l_{M+1}\cdots l_N}
\end{eqnarray}
A familiar special case of (\ref{formula1}) is
$\epsilon^{i_1i_2}_{j_1j_2}=\delta^{i_1}_{j_1}\delta^{i_2}_{j_2}-\delta^{i_1}_{j_2}\delta^{i_2}_{j_1}$.
An example of the use of these identities that is relevant to the correlators computed in this paper is
\begin{eqnarray}\
 \label{identity}
 \epsilon_{i_1\cdots i_N}\epsilon_{k_1\cdots k_N}
\epsilon^{k_1\cdots k_pi_{p+1}\cdots i_N}\epsilon^{i_1\cdots
i_pk_{p+1}\cdots k_N}
&=& \epsilon_{i_1\cdots i_N}^{k_1\cdots k_pi_{p+1}\cdots
i_N}\epsilon_{k_1\cdots k_N}^{i_1\cdots i_pk_{p+1}\cdots k_N}
\nonumber \\
 &=& (N-p)!^2 \epsilon_{i_1\cdots i_p}^{k_1\cdots
k_p}\epsilon_{k_1\cdots k_p}^{i_1\cdots i_p} \nonumber  (N-p)!p!N!
\end{eqnarray}
Another useful summation formula is  ($M\leq N$)
\begin{equation} 
\sum_{p=0}^{M}\frac{(N-p)!}{(M-p)!}=\frac{(N+1)!}{(N-M+1)M!} \, .
\Label{formula4}
\end{equation}
In some computations we only need to keep terms to leading order in large $N$.
Assuming $M$ and $N$ are of the same order (so that $M = \alpha N$ with $0 < \alpha < 1$) it can be shown that at leading order in $N$
\begin{equation}
\sum_{p=0}^{M}\frac{(N-p)!}{(M-p)!}p^k\sim
\sum_{p=0}^{M}\frac{(N-p)!}{(M-p)!}=\frac{(N+1)!}{(N-M+1)M!}\sim\frac{N!}{M!}
\end{equation}
This tells us the factor of $p^k$ only contributes a coefficient of proportionality
at the leading order in large $N$.

\section{Young operators for two column Young diagrams}

In Sec.~ 5 we gave general expressions for the Young operators associated to a Young diagram.  However, it can be difficult in general to construct a concrete closed form expression.  In this appendix we explicitly derive the operators describing two branes with one string on them.  The relevant Young diagrams appear in Fig.~7, and consist of two columns  with $N$ and $M$ boxes ($M \leq N$).   Operators are constructed by filling one box with $Y^J$ and all remaining boxes with $\Phi$.  As discussed in Sec.~5 and Sec.~6, there are two independent states of this form, which we construct below.

\subsection{String attached to the smaller brane}

First, we construct the operator corresponding to the first tableau (1) in Fig.~7, i.e., an open string attached to the smaller brane.    Recall from Sec.~5 that must fill the boxes of the tableau and then symmetrize rows and antisymmetrize columns.   It is easiest to start with the antisymmetrization.   Taking the product of an antisymmetric combination of $N$ $\Phi$s with an antisymmetric combination of $M-1$ $\Phi$s and one $Y^J$ gives the operator
\bea 
P_1 &\equiv &  \epsilon^{i_1 i_2 ... i_N }_{j_1 j_2 ... j_N }
\Phi^{j_1}_{i_1}  \Phi^{j_2}_{i_2}... \Phi^{j_N}_{i_N}
 \epsilon^{\W i_1 \W i_2 ...\W i_M \W k_1 ...\W k_{N-M}}_
{\W j_1 \W j_2... \W j_M \W k_1 ... \W k_{N-M}}
\Phi^{\W j_1}_{\W i_1} \Phi^{\W j_2}_{\W i_2}... 
\Phi^{\W j_{M-1}}_{\W i_{M-1}} (Y^{J})^{\W j_M}_{\W i_M}
\Label{P1}
\eea
It remains to act with the horizontal Young subgroup, i.e., to symmetrize the rows of the tableau.  

To start, we reorganize the sum over $\W i_1$ as
\bean
P_1 & = & \sum_{\W i_1= i_l,l=1}^N
\epsilon^{i_1 i_2 ... i_N }_{j_1 j_2 ... j_N }
\Phi^{j_1}_{i_1}  \Phi^{j_2}_{i_2}... \Phi^{j_N}_{i_N}
 \epsilon^{i_l \W i_2 ...\W i_M \W k_1 ...\W k_{N-M}}_
{\W j_1 \W j_2... \W j_M \W k_1 ... \W k_{N-M}}
\Phi^{\W j_1}_{i_l} \Phi^{\W j_2}_{\W i_2}... 
\Phi^{\W j_{M-1}}_{\W i_{M-1}} (Y^{J})^{\W j_M}_{\W i_M}\\
& = & N \epsilon^{i_1 i_2 ... i_N }_{j_1 j_2 ... j_N }
\Phi^{j_1}_{i_1}  \Phi^{j_2}_{i_2}... \Phi^{j_N}_{i_N}
 \epsilon^{ i_1 \W i_2 ...\W i_M \W k_1 ...\W k_{N-M}}_
{\W j_1 \W j_2... \W j_M \W k_1 ... \W k_{N-M}}
\Phi^{\W j_1}_{ i_1} \Phi^{\W j_2}_{\W i_2}... 
\Phi^{\W j_{M-1}}_{\W i_{M-1}} (Y^{J})^{\W j_M}_{\W i_M} \\
& \equiv & N P_{1,1} \, .
\Label{reorganize1}
\eean
Continuing this way and summing up $\W i_1,...,\W i_k$ gives 
\bea
P_{1} & = & N (N-1) ...(N-k+1) P_{1,k} 
\Label{P1-k} \\
P_{1,k} &=& \epsilon^{i_1 i_2 ... i_N }_{j_1 j_2 ... j_N }
\Phi^{j_1}_{i_1}  \Phi^{j_2}_{i_2}... \Phi^{j_N}_{i_N}
 \epsilon^{ i_1  ...i_k \W i_{k+1}...\W i_M \W k_1 ...\W k_{N-M}}_
{\W j_1 ...\W j_k \W j_{k+1}... \W j_M \W k_1 ... \W k_{N-M}}
\Phi^{\W j_1}_{ i_1}... \Phi^{\W j_k}_{ i_k}
\Phi^{\W j_{k+1}}_{\W i_{k+1}}... 
\Phi^{\W j_{M-1}}_{\W i_{M-1}} (Y^{J})^{\W j_M}_{\W i_M} \ \ \  \ 
\eea

Now consider symmetrizing by action of the horizontal group of the tableau.
Because we have only two columns, all elements of  the horizontal group
are of the form $ (i_{l_1} \W i_{l_1})... (i_{l_k} \W i_{l_k})$
where $(i_{l_1} \W i_{l_1})$ exchanges $i_{l_1}$ and $ \W i_{l_1}$.
A single permutation by 
$(i_1 \W i_1)$, for example, gives
\bean
(i_1 \W i_1)  \cdot P_1 & = &  \epsilon^{i_1 i_2 ... i_N }_{j_1 j_2 ... j_N }
\Phi^{j_1}_{\W i_1}  \Phi^{j_2}_{i_2}... \Phi^{j_N}_{i_N}
 \epsilon^{\W i_1 \W i_2 ...\W i_M \W k_1 ...\W k_{N-M}}_
{\W j_1 \W j_2... \W j_M \W k_1 ... \W k_{N-M}}
\Phi^{\W j_1}_{ i_1} \Phi^{\W j_2}_{\W i_2}... 
\Phi^{\W j_{M-1}}_{\W i_{M-1}} (Y^{J})^{\W j_M}_{\W i_M} 
\Label{singleperm1}
\eean
It is important that we  permute indices before summing over them.
Upon summing over $\W i_1$ on the right hand side, the expression vanishes
unless $\W i_1=i_1$ because of antisymmetry under the $j$ indices.  Thus, 
\bean
(i_1 \W i_1) \cdot  P_1 & = & \epsilon^{i_1 i_2 ... i_N }_{j_1 j_2 ... j_N }
\Phi^{j_1}_{ i_1}  \Phi^{j_2}_{i_2}... \Phi^{j_N}_{i_N}
 \epsilon^{ i_1 \W i_2 ...\W i_M \W k_1 ...\W k_{N-M}}_
{\W j_1 \W j_2... \W j_M \W k_1 ... \W k_{N-M}}
\Phi^{\W j_1}_{ i_1} \Phi^{\W j_2}_{\W i_2}... 
\Phi^{\W j_{M-1}}_{\W i_{M-1}} (Y^{J})^{\W j_M}_{\W i_M} =P_{1,1}
={1\over N} P_1
\eean
Iterating this process, 
\bea 
\prod_{l=1}^k (i_l \W i_l) \cdot  P_1 = k! P_{1,k} =k! { (N-k)!\over N!} P_1
\equiv {1\over C_{N}^{k}} P_1 \Label{k-perm}
\eea

Now it is straight forward to find carry out the symmetrization.  The horizontal
group action is $G = \prod_{l=1}^M (I + (i_l \W i_l))$.   Because (\ref{k-perm}) shows that the action of any product of pair permutations on $P_1$ is proportional to $P_1$ with a coefficient that depends on the number of permutations, we can write $G \cdot P_1 = \left( \sum_{s=0}^M C_M^s P^s \right) \cdot P_1$, where   $P^s$ simply denotes the action of any $s$ pair permutations and
$C_M^s$ is the number of ways of selecting these permutations from the $M$ that appear in the product defining $G$.    Then, using (\ref{k-perm}),
\bea
{\cal O}_{1} & = & 
\prod_{l=1}^M (I + (i_l \W i_l)) \cdot P_1  =  \left( \sum_{s=0}^M C_{M}^s P^s \right) \cdot
P_1 \nonumber \\
& = &  \sum_{s=0}^M C_{M}^s {1\over C_{N}^{s}} P_1
 =  { M!\over N!} P_1 \sum_{s=0}^M { (N-s)!\over (M-s)!} \nonumber \\
& = & { (N+1)\over (N-M+1)} P_1 \Label{Young-1}
\eea
where we have used (\ref{formula4}).

\subsection{String attached to the larger brane}

A string in the larger brane is created by inserting a $Y^J$ is the first  column (which has $N$ boxes) of the two column tableau, i.e., this is the second tableau (2) in Fig.~7.  As above, it is easiest to start with a product operator that has taken care of antisymmetrization by the vertical Young subgroup:
\bea 
P_2= \epsilon^{i_1 i_2 ... i_N }_{j_1 j_2 ... j_N }
\Phi^{j_1}_{i_1}  \Phi^{j_2}_{i_2}... \Phi^{j_{N-1}}_{i_{N-1}}
(Y^{J})^{ j_N}_{ i_N}
 \epsilon^{\W i_1 \W i_2 ...\W i_M \W k_1 ...\W k_{N-M}}_
{\W j_1 \W j_2... \W j_M \W k_1 ... \W k_{N-M}}
\Phi^{\W j_1}_{\W i_1} \Phi^{\W j_2}_{\W i_2}... 
\Phi^{\W j_{M}}_{\W i_{M}} \Label{P2}
\eea
Following the strategy (\ref{reorganize1}) to sum over $\W i_1,..., \W i_k$ gives
\bea \label{k-step-P2}
P_2 & = & { (N-1)!\over (N-k-1)!} P_{2,k,a}+{k (N-1)!\over (N-k)!}
P_{2,k,b}
\eea
where we have defined
\bean \label{P2k-a}
P_{2,k,a} & = & \epsilon^{i_1 i_2 ... i_N }_{j_1 j_2 ... j_N }
\Phi^{j_1}_{i_1}  \Phi^{j_2}_{i_2}... \Phi^{j_{N-1}}_{i_{N-1}}
(Y^{J})^{ j_N}_{ i_N}
\epsilon^{i_1 ...i_k \W i_{k+1} ...\W i_M \W k_1 ...\W k_{N-M}}_
{\W j_1 ...\W j_k \W j_{k+1} ... \W j_M \W k_1 ... \W k_{N-M}}
\Phi^{\W j_1}_{i_1}... \Phi^{\W j_k}_{ i_k}
\Phi^{\W j_{k+1}}_{\W i_{k+1}}... 
\Phi^{\W j_{M}}_{\W i_{M}} \\
P_{2,k,b} & = & \epsilon^{i_1 i_2 ... i_N }_{j_1 j_2 ... j_N }
\Phi^{j_1}_{i_1}  \Phi^{j_2}_{i_2}... \Phi^{j_{N-1}}_{i_{N-1}}
(Y^{J})^{ j_N}_{ i_N}
\epsilon^{i_1 ...i_{k-1} i_N \W i_{k+1} ...\W i_M \W k_1 ...\W k_{N-M}}_
{\W j_1 ...\W j_{k-1} .\W j_k \W j_{k+1} ... \W j_M \W k_1 ... \W k_{N-M}}
\Phi^{\W j_1}_{i_1}... \Phi^{\W j_{k-1}}_{ i_{k-1}}\Phi^{\W j_k}_{ i_N}
\Phi^{\W j_{k+1}}_{\W i_{k+1}}... 
\Phi^{\W j_{M}}_{\W i_{M}} \label{P2k-b}
\eean
The two index structures arise because in reorganizing the sums over $\W i_m$ as sums over the $i_l$ following (\ref{reorganize1}), the index $i_N$ is distinguished because it indexes $Y^J$.   A useful identity that we can infer from the above is
\bean 
P_{2,1,b} & = & { (N-1)!\over (N-k)!} P_{2,k,b}\label{k-step-P2b}  
\eean

With these results in hand,  we symmetrize by the horizontal subgroup.  One permutation gives
\bean
(i_1 \W i_1) \cdot P_2 
& = & P_{2,1,a}+ P_{2,1,c} \\
\label{P21-c}
 P_{2,1,c} & = & \epsilon^{i_1 i_2 ... i_N }_{j_1 j_2 ... j_N }
\Phi^{j_1}_{ i_N}  \Phi^{j_2}_{i_2}... \Phi^{j_{N-1}}_{i_{N-1}}
(Y^{J})^{ j_N}_{ i_N}
 \epsilon^{ i_N \W i_2 ...\W i_M \W k_1 ...\W k_{N-M}}_
{\W j_1 \W j_2... \W j_M \W k_1 ... \W k_{N-M}}
\Phi^{\W j_1}_{ i_1} \Phi^{\W j_2}_{\W i_2}... 
\Phi^{\W j_{M}}_{\W i_{M}} 
\eean
where we permuted $i_i$ and $\W i_1$ as in (\ref{singleperm1}) and then observed that the sum over $\W i_1$ will vanish unless $\W i_1 = i_1$ or $\W i_1 = i_N$.    When applying further permutations to $P_1$ we will find the following definitions and identities useful:
\bea
P_{2,k,c} & = & \epsilon^{i_1 i_2 ... i_N }_{j_1 j_2 ... j_N }
\Phi^{j_1}_{ i_N}  \Phi^{j_2}_{i_2}... \Phi^{j_{N-1}}_{i_{N-1}}
(Y^{J})^{ j_N}_{ i_N}
 \epsilon^{ i_N  i_2... i_k \W i_{k+1} ...\W i_M \W k_1 ...\W k_{N-M}}_
{\W j_1 \W j_2... \W j_k \W j_{k+1}...\W j_M \W k_1 ... \W k_{N-M}}
\Phi^{\W j_1}_{ i_1} \Phi^{\W j_2}_{ i_2}...  \Phi^{\W j_k}_{ i_k}
\Phi^{\W j_{k+1}}_{\W i_{k+1}} ...\Phi^{\W j_{M}}_{\W i_{M}} 
\nonumber  \\
 P_{2,1,c} & = & { (N-2)! \over (N-k-1)!} P_{2,k,c} \\ 
 \label{MX-2} {1\over N} P_{1} & = &
P_{2,1,b}- (N-1) P_{2,1,c} 
\eea
Using these identities we find that 
\begin{eqnarray}
\label{k-perm-P2}
\prod_{l=1}^k (i_l \W i_l) \cdot P_2  =  k! P_{2,k,a}+ k k! P_{2,k,c}
&=& k!  { (N-k-1)!\over (N-1)!} \left[ P_2- k P_{2,1,b}\right] \nonumber \\
& & \ \ \ + \ k k! {(N-k-1)!\over (N-2)!} P_{2,1,c} 
\end{eqnarray}
Finally, we can calculate the Young operator, again using the fact that the action of $s$ permutations on $P_2$ only depends on how many permutations are involved. Using $P^k$ to denote the action of $k$ permutations we find
\bea
\CO_2 = \prod_{l=1}^M (I+(i_l \W i_l)) \cdot P_2  =  \left( \sum_{k=0}^M C_{M}^k P^k \right) \cdot P_2
\Label{permact2}
\eea
Applying our accumulated results gives
\bea 
\CO_2 & = & \sum_{k=0}^M C_{M}^k
\left[  k!  { (N-k-1)!\over (N-1)!} \left[ P_2- k P_{2,1,b}\right]
+k k! {(N-k-1)!\over (N-2)!} P_{2,1,c} \right] \\
& = & {N\over (N-M)} P_2 + { M N(N-1) \over (N-M)(N-M+1)} P_{2,1,c} \nonumber \\
& & \ \ \ - \ { M N \over (N-M)(N-M+1)} P_{2,1,b}
\eea
Finally  using (\ref{MX-2}) we obtain
\bea \label{Young-2-1}
{\cal O}_{2} & = & {N\over (N-M)} P_2 -{ M\over (N-M)(N-M+1)}P_1
\eea
As we will show, this formula can also be derived by demanding orthogonality between the operators making strings on the big and small branes.

\subsection{Orthogonality}

Now we check the orthogonality of $\CO_1$ and $\CO_2$.  Appendix C contains a general proof of orthogonality of operators creating single strings on an arbitrary system of D-branes.

After some tedious but straight forward calculations 
we find that in the planar limit of the free theory
\bea
{\braket{P_1 P_2^\dagger}\over N^{J-1} }&= & 
{ N!^2 (N+1)! (N-1)! (N-M)! M!^2\over (N-M+1) (N-M+2)}
\left[1+ { (J-1) (N-M+1) \over N M}\right]
\Label{P1-P2}\\
{\braket{P_2 P_2^\dagger} \over N^{J-1}} & = & {  N!(N+1)! (N-1)!^2 (N-M)! M!^2 ( (N+1)(N-M)+N)\over
 (N-M+2) (N-M+1)}\nonumber \\ & &  \left[1+ {
(J-1) (N+M) \over N(N+1) ( N(N-M+2)-M) }\right]\Label{P2-P2} \\
{\braket{P_1 P_1^\dagger} \over N^{J-1}} & = & { N!^3 (N+1)! (N-M)! M! (M-1)!  \over (N-M+2)}\left[ 1+ { (J-1) (N-M+1) \over N M}\right]\Label{P1-P1}
\eea
In particular,\footnote{It is worth nothing  that when
$M\to N$, $\braket{P_1 P_1^\dagger}$ does not approach
$\braket{P_2 P_2^\dagger}$ exactly.  Instead,
\bean
{\braket{P_1 P_1^\dagger}-
\braket{P_2 P_2^\dagger}  \over\braket{P_1 P_1^\dagger}+
\braket{P_2 P_2^\dagger} }= { {(J-1)\over N^2}-{2(J-1)\over N(N+1)}
\over 2+{(J-1)\over N^2}+{2(J-1)\over N(N+1)}}
\eean
which is zero in the large 
$N$ limit.  The slight discrepancy occurs because  the contractions of $\epsilon$ tensors are slightly different when $M=N$ (since one of the columns of the Young diagram only has $N-1$ $\Phi$s).  Thus the results above apply strictly to $M < N$.} 
\bea
\braket{P_1 P_2^\dagger} &= & \braket{P_1 P_1^\dagger}
{ M \over N (N-M+1)} \Label{rela-1}
\eea
so 
\bean
\braket{ {\cal O}_{1} {\cal O}_{2}^\dagger} = 0
\eean
which is just what we expected.  Although the explicit results presented above are for planar diagrams it is in fact easy to show that the orthogonality will hold at all orders in the $1/N$ exapansion of the free theory.   The only thing that matters in the calculation is the contraction of the two different index structures of $Y^J$.   A little thought shows that at each non-planar order, these contractions  satisfy (\ref{rela-1}) and  orthogonality of $O_{1,2}$ will follow.

\section{Orthogonality of one open string states}

In this Appendix we demonstrate orthogonality of one open 
string states on arbitrary brane system, i.e, arbitrary
Young diagram. Recall that the operator is given by
\bea
P_{R,p} = \sum_{s\in S_n} \sum_{i=1}^{d_p} D_R(s)_{ii} \cdot s
\Label{Ope-1}
\eea
where $i$ is summing over the irreducible block $p$
of $S_{n-1}$ inside the representaton $R$ of $S_n$. Also we have
used the fact that there is unitary representation matrix with real
number entries so $\O D^R=D^R$. 
Acting with (\ref{Ope-1}) on the fields we have the state
\bea
P_{R,p}(\Phi,Y^J) & = & \sum_{s\in S_n} \sum_{i=1}^{d_p} D_R(s)_{ii}
\sum_{I_i=1}^N \Phi^{I_1}_{I_{s(1)}}  
\Phi^{I_2}_{I_{s(2)}} ... \Phi^{I_{n-1}}_{I_{s({n-1})}} 
(Y^J)^{I_n}_{I_{s(n)}} \Label{Ope-2}\\
P_{R,p}^{\dagger}(\Phi,Y^J) & = & \sum_{t\in S_n} \sum_{i=1}^{d_p} \O D_R(s)_{ii}
\sum_{J_i=1}^N (\Phi^{\dagger})_{J_1}^{J_{t(1)}} 
(\Phi^{\dagger})_{J_2}^{J_{t(2)}} ... 
(\Phi^{\dagger})_{J_{n-1}}^{J_{t({n-1})}}  
(Y^{J\dagger})_{J_n}^{J_{t(n)}}  \Label{Ope-3}
\eea
Note the exchange of up and down indices under
the conjugation of (\ref{Ope-3}).

\subsection{The two point function for $J=1$ case}

Now we calculate the two point fuction as
\bean
I_2& = & \braket{ P_{R,p}(\Phi,Y) P_{S,q}^\dagger(\Phi,Y)} \\
& = & \sum_{I,J} \sum_{s,t\in S_n} \sum_{i,j} D_R(s)_{ii} \O D_S(t)_{jj}
\Phi^{I_1}_{I_{s(1)}}  \Phi^{I_2}_{I_{s(2)}} ... 
\Phi^{I_{n-1}}_{I_{s({n-1})}} 
Y^{I_n}_{I_{s(n)}} 
(\Phi^{\dagger})_{J_1}^{J_{t(1)}} 
(\Phi^{\dagger})_{J_2}^{J_{t(2)}} ... 
(\Phi^{\dagger})_{J_{n-1}}^{J_{t({n-1})}}  
(Y^{\dagger})_{J_n}^{J_{t(n)}}  \\
& = & \sum_{I,J} \sum_{s,t\in S_n} \sum_{i,j} D_R(s)_{ii} \O D_S(t)_{jj}
\sum_{\a\in S_{n-1}} \braket{ \Phi^{I_1}_{I_{s(1)}} 
(\Phi^{\dagger})_{J_{\a(1)}}^{J_{t(\a(1))}} }
\braket{ \Phi^{I_{n-1}}_{I_{s({n-1})}} 
(\Phi^{\dagger})_{J_{\a(n-1)}}^{J_{t(\a(n-1))}}  }
\braket{Y^{I_n}_{I_{s(n)}}  (Y^{\dagger})_{J_n}^{J_{t(n)}}}
\eean
where $\sum_{\a\in S_{n-1}}$ gives all possible contractions.  Now using
\bean
\braket{\Phi^{i}_{j} (\Phi^\dagger)^{k}_l}=\delta^{i}_{l}\delta^{k}_j
{2\pi g_s\over 4\pi^2}{1\over |x-y|^2}
\eean
we get (we have neglected the factor 
${2\pi g_s\over 4\pi^2}{1\over |x-y|^2}$) 
\bean
I_2 & = & \sum_{I,J} \sum_{s,t\in S_n} \sum_{i,j} 
D_R(s)_{ii} \O D_S(t)_{jj} \sum_{\a\in S_{n-1}}
\delta^{I_l}_{J_{\a(l)}} \delta^{J_{t(\a(l))}}_{I_{s(l)}} 
\delta^{I_n}_{J_n}
\delta^{J_{t(n)}}_{I_{s(n)}} \\
& = & \sum_{I,J} \sum_{s,t\in S_n} \sum_{i,j} 
D_R(s)_{ii} \O D_S(t)_{jj} \sum_{\a\in S_{n}|_{\a(n)=n}}
\delta^{I_l}_{J_{\a(l)}} \delta^{J_{t(\a(l))}}_{I_{s(l)}} 
 \\
& = & \sum_{I,J} \sum_{s,t\in S_n} \sum_{i,j} 
D_R(s)_{ii} \O D_S(t)_{jj} \sum_{\a\in S_{n}|_{\a(n)=n}}
\delta^{I_{s(l)}}_{J_{\a(s(l))}} \delta^{J_{t(\a(l))}}_{I_{s(l)}} 
 \\
& = & \sum_{J} \sum_{s,t\in S_n} \sum_{i,j} 
D_R(s)_{ii} \O D_S(t)_{jj} \sum_{\a\in S_{n}|_{\a(n)=n}}
 \delta^{J_{t(\a(l))}}_{J_{\a(s(l))}} 
 \\
& = &  \sum_{s,t\in S_n} \sum_{i,j} 
D_R(s)_{ii} \O D_S(t)_{jj} \sum_{\a\in S_{n}|_{\a(n)=n}}
N^{C(\a^{-1}\cdot t^{-1} \cdot \a \cdot s)}
\eean
where we have used some results from \cite{CJR}. 
Inserting the delta-function 
\bean
I_2 & = & \sum_{s,t\in S_n} \sum_{i,j} 
D_R(s)_{ii} \O D_S(t)_{jj} \sum_{\a\in S_{n}|_{\a(n)=n}}
\sum_{\b\in S_n} N^{C(\b)}\delta(\b^{-1} \cdot 
\a^{-1}\cdot t^{-1} \cdot \a \cdot s) \\
 & = & \sum_{t\in S_n} \sum_{i,j} \sum_{\a\in S_{n}|_{\a(n)=n}}
\sum_{\b\in S_n}
D_R(\a^{-1} t \a \b)_{ii} \O D_S( t)_{jj} 
 N^{C(\b)}\\
 & = & \sum_{\a\in S_{n}|_{\a(n)=n}} \sum_{\b\in S_n} \sum_{i,j}
 N^{C(\b)} 
\sum_{k,l=1}^{d_R} \sum_{t\in S_n}
 D_R(\a^{-1})_{ik} D_R(t)_{kl} D_R(\a\b)_{li}
\O D_S( t)_{jj} 
\eean
Finally we use the orthogonality relation
\bean
\sum_{g\in G} T^{\mu}_{il}(g)\O {T^{\nu}_{sm}}(g)= {d_G\over
d_{\mu}} \delta_{is}\delta_{lm} \delta_{\mu \nu}
\eean
to get
\bea
I_2 & = & \sum_{\a\in S_{n}|_{\a(n)=n}} \sum_{\b\in S_n} \sum_{i,j}
 N^{C(\b)} 
\sum_{k,l} D_R(\a^{-1})_{ik} D_R(\a\b)_{li} {n!\over d_R}
\delta_{RS} \delta_{kj}\delta_{lj} \nonumber \\
 & = & \sum_{\a\in S_{n}|_{\a(n)=n}} \sum_{\b\in S_n} \sum_{i=1}^{d_p}
\sum_{j=1}^{d_q}
 N^{C(\b)} 
 D_R(\a^{-1})_{ij} D_R(\a\b)_{ji} {n!\over d_R}
\delta_{RS} \Label{I2-1}
\eea
From the result (\ref{I2-1})
 we can already read out some orthogonality relationships:

\begin{itemize}

\item (1) If $R\neq S$, $I_2$ is zero. In another words, open
string states are orthogonal if they attach to different brane
systems $R$ and $S$.

\item (2) If $R=S$, but $p\neq q$, $I_2$ is zero. This can be read out 
from $ D_R(\a^{-1})_{ij}$. Since $\a^{-1}(n)=n$, we have
$\a^{-1}\in S_{n-1}\subset S_n$. In the basis we have chosen,
the matrix form is block diagonal so that $ D_R(\a^{-1})_{ij}=0$
if $p\neq q$.
 This proves the
orthogonality of different open string states attaching to
same giant system at tree level.

\item (3) If $R=S$ and $p=q$ we can contiue the calculation as
follows
\bean
I_2 & = & \sum_{\a\in S_{n}|_{\a(n)=n}} \sum_{\b\in S_n} \sum_{i,j}
 N^{C(\b)} 
 D_R(\a^{-1})_{ij} D_R(\a\b)_{ji} {n!\over d_R}
\delta_{RS}\\
& = & \sum_{\a\in S_{n}|_{\a(n)=n}} \sum_{\b\in S_n} \sum_{i}
 N^{C(\b)} D_R(\b)_{ii}{n!\over d_R}
\delta_{RS}\\
& = &{n!(n-1)!\over d_R}\delta_{RS}\sum_{\b\in S_n} N^{C(\b)}
 \sum_{i=1}^{d_p} D_R(\b)_{ii}
\eean

\end{itemize}

\subsection{The two point function with $J>1$}

For $J>1$ a new term shows up in contractions of $Y$ as
\bean
\braket{ (Y^m)^{I_n}_{I_{s(n)}} (Y^{m\dagger})_{J_n}^{J_{t(n)}}} 
\sim N^{m-1} \delta^{I_n}_{J_n} \delta^{J_{t(n)}}_{I_{s(n)}}
+ (m-1) N^{m-2} \delta^{I_n}_{I_{s(n)}} \delta_{J_n}^{J_{t(n)}}
\eean
Thus  we have
\bean
I_2 & = & I_{2,a}+ I_{2,b} \\
I_{2,a} & = & N^{m-1} \sum_{I,J} \sum_{s,t\in S_n} \sum_{i,j} 
D_R(s)_{ii} \O D_S(t)_{jj} \sum_{\a\in S_{n-1}}
 \delta^{I_l}_{J_{\a(l)}} \delta^{J_{t(\a(l))}}_{I_{s(l)}} 
\delta^{I_n}_{J_n}
\delta^{J_{t(n)}}_{I_{s(n)}} \\
I_{2,b} & = & (m-1)N^{m-2} \sum_{I,J} \sum_{s,t\in S_n} \sum_{i,j} 
D_R(s)_{ii} \O D_S(t)_{jj} \sum_{\a\in S_{n-1}}
\delta^{I_l}_{J_{\a(l)}} \delta^{J_{t(\a(l))}}_{I_{s(l)}} 
\delta^{I_n}_{I_{s(n)}}
\delta^{J_{t(n)}}_{J_n} \\
\eean
It is easy to see that $I_{2,a}$ is identical to previous
calculations and we get
\bean
I_{2,a} & = &  N^{m-1}{n!(n-1)!\over d_R} \delta_{RS}\delta_{pq}
\sum_{\b\in S_{n}} N^{C(\b)}  \sum_{i}^{d_p}
 D_R(\b)_{ii} \\
\eean
The remaining hard part is to calculate $I_{2,b}$. 
To do this we need to sum up
\bean
A_2 & = & \sum_{I,J}
\delta^{I_1}_{J_{\a(1)}} \delta^{I_2}_{J_{\a(2)}}...
\delta^{I_{n-1}}_{J_{\a(n-1)}}
\delta^{J_{t(\a(1))}}_{I_{s(1)}} \delta^{J_{t(\a(2))}}_{I_{s(2)}}... 
\delta^{J_{t(\a(n-1))}}_{I_{s(n-1)}} 
\delta^{I_n}_{I_{s(n)}}
\delta^{J_{t(n)}}_{J_n} 
\eean
If we combine
all indices together and treat group elements 
as permutation of $S_{2n}$, we can write $A_2$ as $(I_{s(n)} J_n)\cdot
A_1$ where the permutation $A_1$ is given as
\bean
A_1 & = & \sum_{I,J} \delta^{I_1}_{J_{\a(1)}} \delta^{I_2}_{J_{\a(2)}}...
\delta^{I_{n-1}}_{J_{\a(n-1)}} \delta^{I_n}_{J_n}
\delta^{J_{t(\a(1))}}_{I_{s(1)}} \delta^{J_{t(\a(2))}}_{I_{s(2)}}... 
\delta^{J_{t(\a(n-1))}}_{I_{s(n-1)}} \delta^{J_{t(n)}}_{I_{s(n)}}\\
\eean
Now we can see how the cycle structure changes from $A_1$ to $A_2$
under the permutation $(I_{s(n)} J_n)$:
\begin{itemize}
\item (1)  If cycles of $A_1$ do not
contain $I_{s(n)}$ and $J_n$, they are invariant under the permutation
$(I_{s(n)} J_n)$. 
Because $I_n$ and $J_n$ are always in the same cycle of $S_{2n}$,
when we reduce to the cycle structure of  $S_{n}$, the cycles
of $s\a^{-1} t^{-1} \a$ do not have elements $(n)$ and $s(n)$.

\item (2) If $I_{s(n)}$ and $J_n$ belong to two different
cycles of $A_1$, after the permutation $(I_{s(n)} J_n)$, these
two cycles will combine to one cycle.  
 From the  point of view of $S_{n}$, it means that one cycle
of $s\a^{-1} t^{-1} \a$ has element $(n)$ and another cycle, $s(n)$.
These two cycles will merge by permutation $(n s(n))$ of $S_n$.

\item (3) If $I_{s(n)}$ and $J_n$ belong to same
cycles of $A_1$, after the permutation $(I_{s(n)} J_n)$, this
cycle will break to two cylces. 
 From the  point of view of $S_{n}$, it means that this  cycle
of $s\a^{-1} t^{-1} \a$ has both elements $(n)$ and  $s(n)$ and
it will be broken to two cycle by permutation $(n s(n))$ of $S_n$.

\end{itemize}

All above analyses can be summarized as
\bean
A_2 & = & N^{C(P_s s\a^{-1} t^{-1} \a)}
\eean
where $P_s=(n,s(n))$ is the permutation determined by $s$. 
It is worth to notice that $P_s (s(n))=n$ or $P_s s\in S_{n-1}$.
Using this we can write $\sum_{s\in S_{n}} s\sum_{i=1}^n \sum_{\W s\in S_{n-1}} P_i \W s$ with $P_i=(i,n)$.

Now we can calculate
\bean
I_{2,b} & = & (m-1)N^{m-2} \sum_{\a\in S_{n-1}}
 \sum_{s,t\in S_n} \sum_{i,j}  
D_R(s)_{ii} \O D_S(t)_{jj} 
N^{C(P_s s\a^{-1} t^{-1} \a)}  \\
& = & (m-1)N^{m-2} \sum_{\a\in S_{n-1}}
 \sum_{i=1}^n \sum_{\W s\in S_{n-1}}\sum_{t\in S_n} \sum_{i,j}  
D_R(P_i \W s)_{ii} \O D_S(t)_{jj} 
N^{C(\W  s\a^{-1} t^{-1} \a)}  \\
 & = & (m-1)N^{m-2} \sum_{\a\in S_{n-1}}
 \sum_{k=1}^n \sum_{\W s\in S_{n-1}}\sum_{t\in S_n} \sum_{i,j}  
D_R(P_k \W s)_{ii} \O D_S(t)_{jj} \sum_{\b\in S_n}
N^{C(\b)} \delta(\b^{-1}\W  s\a^{-1} t^{-1} \a)\\
 & = & (m-1)N^{m-2} \sum_{\a\in S_{n-1}}\sum_{\b\in S_n}N^{C(\b)}
 \sum_{k=1}^n \sum_{\W s\in S_{n-1}} \sum_{i,j} 
D_R(P_k \W s)_{ii} \O D_S(\a \b^{-1} \W s \a^{-1})_{jj}\\
 & = & (m-1)N^{m-2} \sum_{\a\in S_{n-1}}\sum_{\b\in S_n}N^{C(\b)}
 \sum_{k=1}^n \sum_{\W s\in S_{n-1}} \sum_{i,j} \\
& & \sum_{l,m}^{d_S}
\sum_{r}^{d_R}
D_R(P_k)_{ir} D_R( \W s)_{ri} \O D_S(\a \b^{-1})_{jl} \O D_S(\W s)_{lm} 
\O D_S( \a^{-1})_{mj}
\eean
Note that from the third equation to the fourth
equation we sum $t\in S_n$ instead of $\W s\in S_{n-1}$ because
the equation $\b^{-1}\W  s\a^{-1} t^{-1} \a=I$ may not have a solution
for all $\W s\in S_{n-1}$. Now we can use the orthogonality relationship
for $\W s\in S_{n-1}$. However, the situation is a little complex
because the representations $R,S$ are irreducible to $S_n$, but
reducible (most time) to $S_{n-1}$. In our choice of bases, the 
matrix is block diagonazed for $\W s\in S_{n-1}$, thus
we have $\sum_{r}^{d_R} D_R( \W s)_{ri}=\sum_{r}^{d_p} D_R( \W s)_{ri}$
where $p$ is the irreducible representation of $S_{n-1}$ inside $R$.
Similarly $\sum_{l,m}^{d_S} \O D_S(\W s)_{lm} =\sum_{q_\gamma} 
\sum_{l_\gamma,m_\gamma}^{d_{\gamma}} \O D_S(\W s)_{l_\gamma m_\gamma}$
where $q_\gamma$ are these irreducible representations of $S_{n-1}$
inside representation $S$. Now using
\bean
\sum_{\W s\in S_{n-1}}D_R( \W s)_{ri}  \O D_S(\W s)_{l_\gamma m_\gamma}
& = & 
{(n-1)!\over d_p} \delta_{p q_\gamma} \delta_{r l_\gamma} \delta_{i  m_\gamma}
\eean we have
\bean
I_{2,b} & = &{(n-1)!(m-1)N^{m-2}\over d_p} \sum_{\a\in S_{n-1}}\sum_{\b\in S_n}N^{C(\b)}
 \sum_{k=1}^n  \sum_{i,j} \\
& & \sum_{r}^{d_p}\sum_{q_\gamma} 
\sum_{l_\gamma,m_\gamma}^{d_{\gamma}}
D_R(P_k)_{ir} \O D_S(\a \b^{-1})_{jl_\gamma} 
\O D_S( \a^{-1})_{m_\gamma j} 
\delta_{p q_\gamma} \delta_{r l_\gamma} \delta_{i  m_\gamma}
\eean
Notice that $\a^{-1}\in S_{n-1}$, $\O D_S( \a^{-1})_{m_\gamma j}$
is not zero when and only when $q_\gamma$ is the same irreducible
representation $q$ chosen by index $j$ (our starting point). Thus we
have 
\bean
I_{2,b} & = &{(n-1)!(m-1)N^{m-2}\over d_p} 
\sum_{\a\in S_{n-1}}\sum_{\b\in S_n}N^{C(\b)}
 \sum_{k=1}^n  \sum_{i,j} \\ & &  \sum_{r}^{d_p} 
\sum_{l,m}^{d_q}
D_R(P_k)_{ir} \O D_S(\a \b^{-1})_{jl} 
\O D_S( \a^{-1})_{m j} 
\delta_{p q} \delta_{r l} \delta_{i  m}\\
 & = &{(n-1)!(m-1)N^{m-2}\over d_p} 
\sum_{\a\in S_{n-1}}\sum_{\b\in S_n}N^{C(\b)}
 \sum_{k=1}^n  \sum_{i} \\ & &  \sum_{r}^{d_p} 
\sum_{l,m}^{d_q} D_R(P_k)_{ir} \O D_S(\a^{-1}\a \b^{-1})_{ml}
\delta_{p q} \delta_{r l} \delta_{i  m}\\
 & = &{(n-1)!(n-1)!(m-1)N^{m-2}\over d_p}
\delta_{p q} \sum_{\b\in S_n}N^{C(\b)}
 \sum_{k=1}^n  \sum_{i}^{d_p} \sum_{l}^{d_q}  
D_{R,p}(P_k \b)_{ii}\\
\eean
Result $I_{2,b}$ tell us  that the orthogonal relationship is not
that perfect because even $R\neq S$, as long as $p\in R$, $q\in S$ and
$p=q$ for $S_{n-1}$  $I_{2,b}$ is still nozero.

Now the large $N$ limit comes to be our rescue.
Let us consider the ratio of ${I_{2,b}\over I_{2,a}}$
\bean
{I_{2,b}\over I_{2,a}} & = & { d_R \over  d_p} { (m-1) \over N n}
{\sum_{\b\in S_n}N^{C(\b)}
 \sum_{k=1}^n  \sum_{i}^{d_p} \sum_{l}^{d_g}  
D_{R,p}(P_k)_{il} D_{S,q}(\b)_{li} \over 
\sum_{\b\in S_{n}} N^{C(\b)}  \sum_{i}^{d_p}
 D_R(\b)_{ii}}
\eean
where ${ d_R \over  d_p}\sim {\cal O}(1)$. The difficult part is
to estimate the last ratio. Notice if $R=S$ 
we have $\sum_{i}^{d_p} \sum_{l}^{d_g}  
D_{R,p}(P_k)_{il} D_{S,q}(\b)_{li}= \sum_{i}^{d_p}D_{R,p}(P_k\b)_{ii}$,
so it is not unreasonable to  guess naively 
that $\sum_{k=1}^n$ will contribute
an order $n$.  If this assumption is right (which we do not know
how to prove it), we have
\bean
{I_{2,b}\over I_{2,a}} \sim {(m-1)\over N}
\eean
Thus as long as the angular momentum $m$ is not large enough, under
proper normalization using $I_{2,a}$ we can neglect the contribution
of $I_{2,b}$ and get the perfect orthogonal relationship, i.e,
{\sl if $R\neq S$ or $p\neq q$, two open string states are orthogonal.}

In principal, we can use similar method to prove the orthogonality for
two open string states or more open string states. However, as
the calculation becomes tedious very quickly, it is better to look
for simpler method.

\section{An example}

In this appendix we will give a detailed example of the correlation function computations in this paper.  Other calculations are carried out similarly, but we omit the details here because of their length.  As our example we will describe how to calculate the two-point functions of the operator $P_1$ (\ref{P11}) which describes two D-branes with a string on the smaller one.  In the free theory the contractions of the fields in (\ref{P11}) give 
\bea
\braket{P_1 P_1^\dagger} 
& = &  (N-M)!^2 \sum_{p=0}^{M-1} C_N^p C_N^p C_{M-1}^p C_{M-1}^p 
p!  p! (N-p)!(M-1-p)!  \nonumber \\
& & \times \ \epsilon^{i_1... i_p i_{p+1} ... i_N}_{j_1 ... j_p j_{p+1} ....
j_N} 
\epsilon^{\W i_1 ... \W i_p \W i_{p+1} ... \W i_{M-1} \W i_M}_{\W j_1 ... \W j_p\W j_{p+1} ... \W j_{M-1} \W j_M}
\epsilon^{\W j_1 ... \W j_p   j_{p+1}...j_N}_{\W i_1 ... \W i_p i_{p+1} ... i_N}
\epsilon^{j_1 ... j_p\W j_{p+1} ... \W j_{M-1} \W k_M}_{i_1 ... i_p
\W i_{p+1} ... \W i_{M-1} \W l_M} \nonumber \\
& & \times \ \left[ N^{J-1} \delta^{\W j_M}_{\W k_M}\delta_{\W i_M}^{\W l_M}
+(J-1) N^{J-2}  \delta^{\W j_M}_{\W i_M} \delta_{\W k_M}^{\W l_M}
\right]
\eea
Here $p$ counts the number of $\Phi$ fields in the big brane of $P_1$ that are contracted with $\bar\Phi$ fields in the small brane of $P_1^\dagger$.   The combinatorial factor $ C_N^p C_N^p C_{M-1}^p C_{M-1}^p  p!  p! (N-p)!(M-1-p)!$ arises from the number of ways in which these contractions can be done.  For example, choosing $p$ $\Phi$ from the large brane
in $P_1$ and $p$ $\O \Phi$ from  
small brane of $\O P_1$ gives factor $ C_N^p C_{M-1}^p$
while different orders of contractions among these two groups of $\Phi$
gives the factor $p!$.  The contractions between $Y$s give two
different index structures.   While the contraction between the $\Phi$s are carried out exactly, those between the $Y$s are done at the planar level.  In this sense our computations are at the leading order in large $N$.   One loop corrections from interactions are also evaluated in the planar limit for contractions between $Y$s.

For the first index structure we have
\bea
\braket{P_1 P_1^\dagger}_a & = & 
(N-M)!^2 N^{J-1}\sum_{p=0}^{M-1} C_N^p C_N^p C_{M-1}^p C_{M-1}^p 
p!  p! (N-p)!(M-1-p)!  \nonumber \\
& & \times \ \epsilon^{i_1... i_p i_{p+1} ... i_N}_{j_1 ... j_p j_{p+1} ....
j_N} 
\epsilon^{\W i_1 ... \W i_p \W i_{p+1} ... \W i_{M-1} \W i_M}_{\W j_1 ... \W j_p\W j_{p+1} ... \W j_{M-1} \W j_M}
\epsilon^{\W j_1 ... \W j_p   j_{p+1}...j_N}_{\W i_1 ... \W i_p i_{p+1} ... i_N}
\epsilon^{j_1 ... j_p\W j_{p+1} ... \W j_{M-1} \W j_M}_{i_1 ... i_p
\W i_{p+1} ... \W i_{M-1} \W i_M} 
\eea
We can interchange the lower indices of first and third $\epsilon$ tensors, because both of them are maximal to get 
\bea
\braket{P_1 P_1^\dagger}_a
& = & 
(N-M)!^2 N^{J-1}\sum_{p=0}^{M-1} C_N^p C_N^p C_{M-1}^p C_{M-1}^p 
p!  p! (N-p)!(M-1-p)!  \nonumber \\
& & \times \  \epsilon^{i_1... i_p i_{p+1} ... i_N}_{\W i_1 ... \W i_p i_{p+1} ... i_N}
\epsilon^{\W i_1 ... \W i_p \W i_{p+1} ... \W i_{M-1} \W i_M}_{\W j_1 ... \W j_p\W j_{p+1} ... \W j_{M-1} \W j_M}
\epsilon^{\W j_1 ... \W j_p   j_{p+1}...j_N}_{j_1 ... j_p j_{p+1} ....
j_N}
\epsilon^{j_1 ... j_p\W j_{p+1} ... \W j_{M-1} \W j_M}_{i_1 ... i_p
\W i_{p+1} ... \W i_{M-1} \W i_M}
\eea
Carrying out various index contractions using the identities in Appendix A gives
\bea
\braket{P_1 P_1^\dagger}_a
& = & 
N-M)!^2 N^{J-1}\sum_{p=0}^{M-1} C_N^p C_N^p C_{M-1}^p C_{M-1}^p 
p!  p! (N-p)!(M-1-p)! \nonumber \\
& & \times \ (N-p)!^2 \epsilon^{i_1... i_p }_{\W i_1 ... \W i_p }
\epsilon^{\W i_1 ... \W i_p \W i_{p+1} ... \W i_{M-1} \W i_M}_{\W j_1 ... \W j_p\W j_{p+1} ... \W j_{M-1} \W j_M}
\epsilon^{\W j_1 ... \W j_p   }_{j_1 ... j_p }
\epsilon^{j_1 ... j_p\W j_{p+1} ... \W j_{M-1} \W j_M}_{i_1 ... i_p
\W i_{p+1} ... \W i_{M-1} \W i_M} \\
& = & 
(N-M)!^2 N^{J-1}\sum_{p=0}^{M-1} C_N^p C_N^p C_{M-1}^p C_{M-1}^p 
p!  p! (N-p)!(M-1-p)! \nonumber \\
& & \times \  (N-p)!^2 p!^2
\epsilon^{ i_1 ...  i_p \W i_{p+1} ... \W i_{M-1} \W i_M}
_{\W j_1 ... \W j_p\W j_{p+1} ... \W j_{M-1} \W j_M}
\epsilon^{\W j_1 ... \W j_p\W j_{p+1} ... \W j_{M-1} \W j_M}_{i_1 ... i_p
\W i_{p+1} ... \W i_{M-1} \W i_M}
\Label{fourthequation}
\eea
Further simplification leads to
\bea
\braket{P_1 P_1^\dagger}_a
& = & 
(N-M)!^2 N^{J-1}\sum_{p=0}^{M-1} C_N^p C_N^p C_{M-1}^p C_{M-1}^p 
p!  p! (N-p)!(M-1-p)! \nonumber \\
& & \times \ (N-p)!^2 p!^2  M! { N!\over (N-M)!}\\
& = & N^{J-1} N!^3 (N-M)! M! (M-1)!^2 \sum_{p=0}^{M-1} { (N-p)!\over (M-1-p)!}\\
& = & N^{J-1} N!^3 (N-M)! M! (M-1)!^2 { (N+1)! \over (N-M+2) (M-1)!}
\eea
In this particular case the final index contractions in (\ref{fourthequation}) were simple. However, the $\braket{P_1 P_1^\dagger}$ has two
index structures. If we evaluate the second index structure 
$(J-1) N^{J-2}  \delta^{\W j_M}_{\W i_M} \delta_{\W k_M}^{\W l_M}$
we will meet following index contractions
\bean
{\cal E}_{II}& \equiv&
\epsilon^{j_p j_{N-p-1} j}_{i_p i_{N-p-1} j}
\O \epsilon^{k_p i_{N-p-1} i}_{l_p j_{N-p-1} i}
\epsilon^{l_p l_{M-p}}_{k_p k_{M-p}}
\O \epsilon^{i_p k_{M-p}}_{j_p l_{M-p}} 
 =  \epsilon^{j_p j_{N-p-1} j}_{l_p j_{N-p-1} i}
\O \epsilon^{k_p i_{N-p-1} i}_{i_p i_{N-p-1} j}
\epsilon^{l_p l_{M-p}}_{k_p k_{M-p}}
\O \epsilon^{i_p k_{M-p}}_{j_p l_{M-p}}\\
& = & (N-p-1)!^2 \epsilon^{j_p j}_{l_p i}
\O \epsilon^{k_p  i}_{i_p j}
\epsilon^{l_p l_{M-p}}_{k_p k_{M-p}}
\O \epsilon^{i_p k_{M-p}}_{j_p l_{M-p}}
\eean
where we have used $j_p$ to represent $j_1...j_p$ and $k_{M-p}$ to
represent $k_1...k_{M-p}$ to make notation simpler. 
The factor $\epsilon^{j_p j}_{l_p i}
\O \epsilon^{k_p  i}_{i_p j}
\epsilon^{l_p l_{M-p}}_{k_p k_{M-p}}
\O \epsilon^{i_p k_{M-p}}_{j_p l_{M-p}}$ can be computed step by
step:

\begin{itemize}

\item (1) Fixing $i$ we have $N$ choices.

\item (2) If $j=i$, then $i_p, j_p, k_p, l_p$ can only choose
from remaining $(N-1)$ values and we denote as
$i_p, j_p, k_p, l_p \in U(N-1)$. Now
 $l_{M-p}, k_{M-p}$ can have two realization. The first
 is that one index in the set $l_{M-p}, k_{M-p}$ (recalling
that these two sets have $(M-p)$ indices) take the value $i$.
The second is that no index in the set $l_{M-p}, k_{M-p}$  
take the value $i$.
For the first case we can write $l_{M-p}= i l_{M-p-1}$ and
$k_{M-p}= i k_{M-p-1}$ after some permutations. Thus we have 
\bean
& & N (M-p)^2 \epsilon^{j_p}_{l_p}\O \epsilon^{k_p }_{i_p}
\epsilon^{l_p l_{M-p-1}}_{k_p k_{M-p-1}}
\O \epsilon^{i_p k_{M-p-1}}_{j_p l_{M-p-1}}\\
& = &  N (M-p)^2 p!^2 \epsilon^{j_p l_{M-p-1}}_{k_p k_{M-p-1}}
\O \epsilon^{k_p k_{M-p-1}}_{j_p l_{M-p-1}}\\
& = &  N (M-p)^2 p!^2 (M-1)! { (N-1)! \over (N-1-(M-1))!} \\
& = & { N! (M-1)! p!^2 (M-p)^2\over (N-M)!}
\eean
The factor $N$ comes from summing $i$. The factor $(M-p)^2$
comes from the choice of one index from two sets $l_{M-p}, k_{M-p}$.
Furthermore, when writting in above form we have neglected
the index $i$. The purpose of doing so is because now every
index can only take values from remaining $(N-1)$ values. In another
word, we can treat above expression as the index contraction
for $U(N-1)$ gauge group (recalling  results in the Appendix A).

For the second case we have 
\bean
& & N  \epsilon^{j_p}_{l_p}\O \epsilon^{k_p }_{i_p}
\epsilon^{l_p l_{M-p}}_{k_p k_{M-p}}
\O \epsilon^{i_p k_{M-p-1}}_{j_p l_{M-p}}\\
& = & N  p!^2 \epsilon^{j_p l_{M-p}}_{k_p k_{M-p}}
\O \epsilon^{k_p k_{M-p}}_{j_p l_{M-p}}\\
& = &  N  p!^2  M! { (N-1)! \over (N-1-M)!}  \\
& = & { N! (M-1)! p!^2  M(N-M)\over (N-M)!}
\eean
Adding these two cases together we have 
\bean
{ N! (M-1)! p!^2 \over (N-M)!} [(M-p)^2+ M(N-M)]
\eean

\item (3). Now we discuss the case that $i\neq j$. It is obvious
that we will have factor $N(N-1)$ from summing $i,j$. Since
 $\epsilon^{j_p j}_{l_p i}$ is zero unless $(l_p i)$ is a permutation
of $(j_p j)$, this means that index $j$ must show up in set $l_p$. 
Similarly $i$ must show up in set $j_p$. After all we have
  $i\in j_p, i_p$, $j\in l_p, k_p$ with combinatorical factor $p^4$. 
Notice that $l_{M-p}, k_{M-p}\in U(N-2)$
so we reduced index contraction to
\bean
& & N(N-1) p^4 \epsilon^{j_{p-1}}_{l_{p-1}}
\O \epsilon^{k_{p-1} }_{i_{p-1}}
\epsilon^{l_{p-1} l_{M-p}}_{k_{p-1} k_{M-p}}
\O \epsilon^{i_{p-1} k_{M-p-1}}_{j_{p-1} l_{M-p}}\\
& = & N(N-1) p^4 (p-1)!^2
\epsilon^{j_{p-1} l_{M-p}}_{i_{p-1} k_{M-p}}
\O \epsilon^{i_{p-1} k_{M-p-1}}_{j_{p-1} l_{M-p}}\\
& = & N(N-1) p^4 (p-1)!^2 (M-1)! { (N-2)! \over (N-M-1)!}\\
& = & { N! (M-1)! p!^2 \over (N-M)!} [ p^2 (N-M)]
\eean
where indices $i,j$ have been deleted and contraction
is done in effective $U(N-2)$ gauge theory.

\item (4) Adding (2) and (3) together we have 
\bean
 \epsilon^{j_p j}_{l_p i}
\O \epsilon^{k_p  i}_{i_p j}
\epsilon^{l_p l_{M-p}}_{k_p k_{M-p}}
\O \epsilon^{i_p k_{M-p}}_{j_p l_{M-p}}={ N! (M-1)! p!^2 \over (N-M)!}[(M-p)^2+ M(N-M)+ p^2 (N-M)]
\eean

\end{itemize}

Putting factor $(N-p-1)!^2$ back we finally have
\bean
 {\cal E}_{II} & = & { N! (N-p-1)!^2 p!^2  M!\over (N-M-1)!}
\left[ 1+{(M-p)^2 \over
M(N-M)}+ {p^2\over M}\right]\\
& = & { N! (M-1)! p!^2 (N-p-1)!^2
 \over (N-M)!}[(M-p)^2+ M(N-M)+ p^2 (N-M)]
\eean
To calculate the complete one-loop correction we need to insert four
interaction terms. For each term, there are in principal $4!=24$ possible index contractions where each one could be as (pr more) complicated
as (than) ${\cal E}_{II}$. From this we see that 
 an explicit calculation of the  one-loop correction 
is rather complicated.
Thus our results provide a non-trivial check of our proposal concerning emergent gauge symmetry from D-brane operators in Yang-Mills theory.


\begin{thebibliography}{10}
\newcommand{\wwwspires}{http://www.slac.stanford.edu/spires/find/hep/www}



\bbibitem{thooft}
G. 't Hooft, "A Planar Diagram Theory for Strong Interactions,"
{\em Nucl. Phys.}{\bf B 72} 461 (1974)




\bbibitem{jthroat}
J.~Maldacena, ``The Large N limit of superconformal field theories
and supergravity,'' {\em Adv. Theor. Math. Phys.} {\bf 2} (1998)
231, {{\tt hep-th/9711200}};~~~S.~S.~Gubser, I.~R.~Klebanov, and
A.~M.~Polyakov, ``Gauge theory correlators from noncritical string
theory,'' {\em Phys. Lett.} {\bf B428} (1998) 105,
 {{\tt hep-th/9802109}};~~~
E.~Witten, ``Anti-de Sitter space and holography,'' {\em Adv.
Theor. Math. Phys.} {\bf 2} (1998) 253, {{\tt hep-th/9802150}}.

\bbibitem{MST}
J. McGreevy, L. Susskind and N. Toumbas, ``Invasion of the giant
gravitons from anti-de Sitter space,'' {\em JHEP} {\bf 0006}
(2000) 008 {\tt hep-th/0003075}

\bbibitem{akietal}
 A.~Hashimoto, S.~Hirano and
N.~Itzhaki, ``Large branes in AdS and their field theory dual,''
JHEP {\bf 0008}, 051 (2000) [arXiv:hep-th/0008016]

\bbibitem{goliath}
M.~T.~Grisaru, R.~C.~Myers and O.~Tafjord, ``SUSY and Goliath,''
JHEP {\bf 0008}, 040 (2000) [arXiv:hep-th/0008015].

\bbibitem{BBNS}
V. Balasubramanian, M. Berkooz, A. Naqvi and M. Strassler, ``Giant
gravitons in conformal field theory,'' {\tt hep-th/0107119}.

\bbibitem{CJR}
S. Corley, A. Jevicki and S. Rangoolam, ``Exact correlators of
giant gravitons from dual N = 4 SYM theory" , {\tt
hep-th/0111222}.

\bbibitem{Btoy}
D.~Berenstein,
``A toy model for the AdS/CFT correspondence,''
JHEP {\bf 0407}, 018 (2004)
[arXiv:hep-th/0403110].

\bbibitem{linluninmalda}
H.~Lin, O.~Lunin and J.~Maldacena,
``Bubbling AdS space and 1/2 BPS geometries,''
arXiv:hep-th/0409174.

\bibitem{Calda}
M.~M.~Caldarelli and P.~J.~Silva,
``Giant gravitons in AdS/CFT. I: Matrix model and back reaction,''
JHEP {\bf 0408}, 029 (2004)
[arXiv:hep-th/0406096].




\bbibitem{BHLN}
V.~Balasubramanian, M.~x.~Huang, T.~S.~Levi and A.~Naqvi, ``Open
strings from N = 4 super Yang-Mills,'' JHEP {\bf 0208}, 037 (2002)
[arXiv:hep-th/0204196].


\bbibitem{Bgiant}
D.~Berenstein,
``Shape and holography: Studies of dual operators to giant gravitons,''
Nucl.\ Phys.\ B {\bf 675}, 179 (2003)
[arXiv:hep-th/0306090].

\bbibitem{AABF}
O.~Aharony, Y.~E.~Antebi, M.~Berkooz and R.~Fishman,
``'Holey sheets': Pfaffians and subdeterminants as D-brane operators in large
N gauge theories,''
JHEP {\bf 0212}, 069 (2002)
[arXiv:hep-th/0211152].


\bibitem{ShJ}
M.~M.~Sheikh-Jabbari,
``Tiny graviton matrix theory: DLCQ of IIB plane-wave string theory, a
conjecture,''
JHEP {\bf 0409}, 017 (2004)
[arXiv:hep-th/0406214].




\bibitem{sadri}
D.~Sadri and M.~M.~Sheikh-Jabbari,
``Giant hedge-hogs: Spikes on giant gravitons,''
Nucl.\ Phys.\ B {\bf 687}, 161 (2004)
[arXiv:hep-th/0312155].


\bbibitem{matrix}
J.~Polchinski,
``Dirichlet-Branes and Ramond-Ramond Charges,''
Phys.\ Rev.\ Lett.\  {\bf 75}, 4724 (1995)
[arXiv:hep-th/9510017].
E.~Witten,
``Bound states of strings and p-branes,''
Nucl.\ Phys.\ B {\bf 460}, 335 (1996)
[arXiv:hep-th/9510135].

\bbibitem{Das}
S.~R.~Das, A.~Jevicki and S.~D.~Mathur, ``Vibration modes of giant
gravitons,'' Phys.\ Rev.\ D {\bf 63}, 024013 (2001)
[arXiv:hep-th/0009019].



\bbibitem{BMN}
D.~Berenstein, J.~M.~Maldacena and H.~Nastase,
``Strings in flat space and pp waves from N = 4 super Yang Mills,''
JHEP {\bf 0204}, 013 (2002)
[arXiv:hep-th/0202021].

\bbibitem{wilson}
S.~J.~Rey and J.~T.~Yee,
``Macroscopic strings as heavy quarks in large N gauge theory and  anti-de
Sitter supergravity,''
Eur.\ Phys.\ J.\ C {\bf 22}, 379 (2001)
[arXiv:hep-th/9803001].
J.~M.~Maldacena,
``Wilson loops in large N field theories,''
Phys.\ Rev.\ Lett.\  {\bf 80}, 4859 (1998)
[arXiv:hep-th/9803002].




 \bbibitem{BHK}
D.~Berenstein, C.~P.~Herzog and I.~R.~Klebanov,
``Baryon spectra and AdS/CFT correspondence,''
JHEP {\bf 0206}, 047 (2002)
[arXiv:hep-th/0202150].

\bbibitem{Constable}
N.~R.~Constable, D.~Z.~Freedman, M.~Headrick, S.~Minwalla,
L.~Motl, A.~Postnikov and W.~Skiba, ``PP-wave string interactions
from perturbative Yang-Mills theory,'' JHEP {\bf 0207}, 017 (2002)
[arXiv:hep-th/0205089].

\bbibitem{DHoker}
E.~D'Hoker, D.~Z.~Freedman and W.~Skiba, ``Field theory tests for
correlators in the AdS/CFT correspondence,'' Phys.\ Rev.\ D {\bf
59}, 045008 (1999) [arXiv:hep-th/9807098].

\bbibitem{Skiba}
W.~Skiba, ``Correlators of short multi-trace operators in N = 4
supersymmetric  Yang-Mills,'' Phys.\ Rev.\ D {\bf 60}, 105038
(1999) [arXiv:hep-th/9907088].


\bbibitem{parklee}
P.~Lee and J.~w.~Park,
``Open strings in PP-wave background from defect conformal field theory,''
Phys.\ Rev.\ D {\bf 67}, 026002 (2003)
[arXiv:hep-th/0203257].


\bbibitem{DP}
A.~Dabholkar and S.~Parvizi,
``Dp branes in pp-wave background,''
Nucl.\ Phys.\ B {\bf 641}, 223 (2002)
[arXiv:hep-th/0203231].


\bbibitem{kostasmarika}
K.~Skenderis and M.~Taylor,
``Branes in AdS and pp-wave spacetimes,''
JHEP {\bf 0206}, 025 (2002)
[arXiv:hep-th/0204054].

\bbibitem{openstrings}
D.~Berenstein, E.~Gava, J.~M.~Maldacena, K.~S.~Narain and H.~Nastase,
``Open strings on plane waves and their Yang-Mills duals,''
arXiv:hep-th/0203249.


\bibitem{BQHE}
D.~Berenstein,
``A matrix model for a quantum Hall droplet with manifest particle-hole
symmetry,''
arXiv:hep-th/0409115.


\bibitem{DeM}
R.~de Mello Koch and R.~Gwyn,
``Giant graviton correlators from dual SU(N) super Yang-Mills theory,''
arXiv:hep-th/0410236.

\bbibitem{groupbook}
W.~Miller Jr., ``Symmetry groups and their applications'', Academic Press, New York, 1972.


\bbibitem{groupbook2}
W.~Fulton and J.~Harris, ``Representation theory'',  Springer, New York, 1991.


\bbibitem{Huang}
M.~x.~Huang, ``Three point functions of N = 4 super Yang Mills
from light cone string field theory in pp-wave,'' Phys.\ Lett.\ B
{\bf 542}, 255 (2002) [arXiv:hep-th/0205311];  M.~x.~Huang,
``String interactions in pp-wave from N = 4 super Yang Mills,''
Phys.\ Rev.\ D {\bf 66}, 105002 (2002) [arXiv:hep-th/0206248].

\bbibitem{SV}
M.~Spradlin and A.~Volovich, ``Superstring interactions in a
pp-wave background,'' Phys.\ Rev.\ D {\bf 66}, 086004 (2002)
[arXiv:hep-th/0204146];  M.~Spradlin and A.~Volovich,
``Superstring interactions in a pp-wave background. II,'' JHEP
{\bf 0301}, 036 (2003) [arXiv:hep-th/0206073].

\bbibitem{Open}
B.~Chandrasekhar and A.~Kumar, ``D-branes in pp-wave light cone
string field theory,'' JHEP {\bf 0306}, 001 (2003)
[arXiv:hep-th/0303223];  B.~.~J.~Stefanski, ``Open string
plane-wave light-cone superstring field theory,''
arXiv:hep-th/0304114;  J.~Lucietti, S.~Schafer-Nameki and
A.~Sinha, ``On the exact open-closed vertex in plane-wave
light-cone string field theory,'' arXiv:hep-th/0311231; J.~Gomis,
S.~Moriyama and J.~w.~Park, ``Open + closed string field theory
from gauge fields,'' arXiv:hep-th/0305264.


\bbibitem{vijayigor}
V.~Balasubramanian and I.~R.~Klebanov,
``Some Aspects of Massive World-Brane Dynamics,''
Mod.\ Phys.\ Lett.\ A {\bf 11}, 2271 (1996)
[arXiv:hep-th/9605174].

\bbibitem{iengorusso}
R.~Iengo and J.~G.~Russo,
``The decay of massive closed superstrings with maximum angular momentum,''
JHEP {\bf 0211}, 045 (2002)
[arXiv:hep-th/0210245];~~~
R.~Iengo and J.~G.~Russo,
``Semiclassical decay of strings with maximum angular momentum,''
JHEP {\bf 0303}, 030 (2003)
[arXiv:hep-th/0301109].


\bbibitem{BN}
V.~Balasubramanian and A.~Naqvi, ``Giant gravitons and a
correspondence principle,'' Phys.\ Lett.\ B {\bf 528}, 111 (2002)
[arXiv:hep-th/0111163].


\bbibitem{Gub}
S.~S.~Gubser and J.~J.~Heckman,
``Thermodynamics of R-charged black holes in AdS(5) from effective strings,''
arXiv:hep-th/0411001.


\bbibitem{nemani}
N.~V.~Suryanarayana,
``Half-BPS Giants, Free Fermions and Microstates of Superstars,''
arXiv:hep-th/0411145.



\bibitem{BDHM}
T.~Banks, M.~R.~Douglas, G.~T.~Horowitz and E.~J.~Martinec,
``AdS dynamics from conformal field theory,''
arXiv:hep-th/9808016.


\bibitem{BKLT}
V.~Balasubramanian, P.~Kraus, A.~E.~Lawrence and S.~P.~Trivedi,
``Holographic probes of anti-de Sitter space-times,''
Phys.\ Rev.\ D {\bf 59}, 104021 (1999)
[arXiv:hep-th/9808017].

\bbibitem{Gidd}
S.~B.~Giddings,
``Flat-space scattering and bulk locality in the AdS/CFT  correspondence,''
Phys.\ Rev.\ D {\bf 61}, 106008 (2000)
[arXiv:hep-th/9907129].


\bibitem{HH}
G.~T.~Horowitz and V.~E.~Hubeny,
``CFT description of small objects in AdS,''
JHEP {\bf 0010}, 027 (2000)
[arXiv:hep-th/0009051].


\bibitem{PS}
J.~Polchinski and M.~J.~Strassler,
``Hard scattering and gauge/string duality,''
Phys.\ Rev.\ Lett.\  {\bf 88}, 031601 (2002)
[arXiv:hep-th/0109174].

















\end{thebibliography}
\end{document}